\documentclass[journal ]{new-aiaa}
\usepackage[utf8]{inputenc}
\usepackage{textcomp}
\usepackage{graphicx}
\usepackage{amsmath}
\usepackage[version=4]{mhchem}
\usepackage{siunitx}
\usepackage{longtable,tabularx}
\usepackage{float, graphicx}
\usepackage[capitalise]{cleveref}
\usepackage[ruled,vlined,linesnumbered]{algorithm2e}
\SetKwComment{Comment}{/* }{ */}
\usepackage{color,xcolor}
\renewcommand{\deg}{^\circ}
\newcommand{\piPAG}{$\pi$PAG~}
\newcommand{\eFNPAG}{$\varepsilon$FNPAG~}
\usepackage{tikz}
\definecolor{layerfill}{HTML}{CFB87C}
\usepackage{caption}
\usepackage{subcaption}
\usepackage{tikz}
\usetikzlibrary{automata,arrows,positioning,calc}
\usepackage{soul}
\usepackage{multirow}
\usepackage{makecell}

\title{Risk-Aware Aerocapture Guidance Through a Probabilistic Indicator Function}

\author{Grace E. Calkins,\footnote{Ph.D. Student, Department of Aerospace Engineering Sciences. grace.calkins@colorado.edu} Jay W. McMahon,\footnote{Associate Professor, Department of Aerospace Engineering Sciences.} and Alireza Doostan\footnote{Professor, Department of Aerospace Engineering Sciences.}}
\affil{University of Colorado Boulder, Boulder, CO 80303}
\author{David C. Woffinden\footnote{Aerospace Engineer, GN\&C Autonomous Systems Branch}}
\affil{NASA Johnson Space Center, Houston, TX, 77508}

\begin{document}

\maketitle

\begin{abstract}
    Aerocapture is sensitive to trajectory errors, particularly for low-cost missions with imprecise navigation. For such missions, considering the probability of each failure mode when computing guidance commands can increase capture rate. A risk-aware aerocapture guidance algorithm is proposed that uses a generative model-based probabilistic indicator function to estimate escape, impact, or capture probabilities. The probability of each mode is incorporated into corrective guidance commands to increase the likelihood of successful capture. The proposed method is evaluated against state-of-the-art numeric predictor-corrector guidance algorithms in high-uncertainty scenarios where entry interface dispersions lead to nontrivial failure probabilities. When using a probabilistic indicator function in guidance, 71.43\% to 100\% of recoverable cases are saved for a variety of initial distributions and atmosphere models. The probabilistic indicator function is capable of predicting failure probability for dispersions and atmosphere models outside its training data, showing generalizability. In addition, the probabilistic indicator is compared to a fading memory filter for density estimation, demonstrating improvements in accuracy when both are used in conjunction. The proposed risk-aware aerocapture guidance algorithm improves capture performance and robustness to entry interface state dispersions, especially for missions with high navigation uncertainty.
\end{abstract}

\section*{Nomenclature}

{\renewcommand\arraystretch{1.0}
\noindent
\begin{longtable*}{@{}l @{\quad=\quad} l@{}}
$A$ & Downsampling cluster weight strength \\
$A_{\text{ref}}$ & Reference area [m$^2$] \\
$a$ & Semimajor axis [m] \\
$C$ & Number of Gaussian mixture clusters \\
$C_D$ & Drag coefficient \\
$C_L$ & Lift coefficient \\
$c_1, \ldots, c_9$ & Polynomial atmosphere model coefficients \\
$d$ & Gaussian mixture variational autoencoder latent dimension \\
$d^h_i$ & Gaussian mixture variational autoencoder network hidden dimension \\
$e$ & Orbit eccentricity \\ 
$\bar{e}_{\text{W}}$ & Weighted failure rate \\ 
$\bar{e}_{\text{F}}$ & Failure-only weighted failure rate \\ 
$F_N$ & Normal (lift) acceleration [m/s$^2$] \\
$F_T$ & Tangential (drag) acceleration [m/s$^2$] \\
$g_0$ & Gravity at planet's surface [m/s$^2$] \\
$g_\phi$ & Latitudinal component of gravity [m/s$^2$] \\
$g_r$ & Radial component of gravity [m/s$^2$] \\
$h$ & Altitude above planetary surface [m] \\
$J_2$ & Second zonal harmonic of the planet \\
$k_{i,\text{low}}$ & Gaussian-Uniform mixture lower tail bound coefficient \\
$k_{i,\text{high}}$ & Gaussian-Uniform mixture upper tail bound coefficient \\
$l$ & Sample cluster label \\
$\frac{L}{D}$ & Lift-to-drag ratio \\
$m$ & Vehicle mass [kg] \\
$n$ & Gaussian mixture variational autoencoder input dimension \\
$P_{\text{outcome}}$ & Probability of outcome \\
$q$ & Variational posterior distribution \\
$R_e$ & Equatorial radius of the planet [m] \\
$r$ & Radius from planet center [m] \\
$r_a$ & Apoapsis radius [m] \\
$r_p$ & Periapsis radius [m] \\
$t_c$ & Downsampling center time \\ 
$V$ & Planet-relative velocity [m/s] \\
$w$ & Downsampling weight  \\
$\mathbf{x}$ & Gaussian mixture variational autoencoder input \\ 
$\mathbf{z}$ & Gaussian mixture variational autoencoder latent encoding \\ 
$\beta$ & Ballistic coefficient [kg/m$^2$] \\
$\beta_{\text{KL}}$ & Variational autoencoder Kullback-Leibler divergence weighting factor \\
$\gamma$ & Flight path angle [rad] \\
$\gamma_{ic}$ & Gaussian mixture variational autoencoder cluster membership probability  \\
$\Delta i$ & Change in inclination \\
$\Delta v$ & Propellant required \\
$\varepsilon$ & Specific energy [m$^2$/s$^2$] \\
$\epsilon_C$ & Capture probability threshold \\
$\epsilon_F$ & Failure probability threshold \\
$\theta$ & Longitude [$^\circ$] \\
$\zeta$ & Bank angle time constant [s] \\
$\kappa$ & Gaussian mixture variational autoencoder encoder parameters \\
$\lambda$ & Gaussian mixture variational autoencoder decoder parameters \\
$\mu$ & Gravitational parameter of the planet [m$^3$/s$^2$] \\
$\boldsymbol{\mu}_c$ & Gaussian mixture variational autoencoder cluster means \\
$\nu$ & Student's $t$ distribution degree of freedom \\
$\pi_c$ & Gaussian mixture variational autoencoder cluster weight \\
$\rho$ & Atmospheric density [kg/m$^3$] \\
$\rho_D$ & Drag fading filter scaling factor \\
$\rho_L$ & Lift fading filter scaling factor \\
$\tau$ & Bank angle correction persistence time [s] \\
$\phi$ & Latitude [$^\circ$] \\
$\phi_x$ & Gaussian mixture variational autoencoder ELBO recognition network \\
$\phi_c$ & Gaussian mixture variational autoencoder ELBO recognition network \\
$\chi$ & Fading filter time constant  \\
$\psi$ & Heading angle [$^\circ$] \\
$\sigma$ & Bank angle [$^\circ$] \\
$\sigma^*$ & Commanded bank angle [$^\circ$] \\
$\sigma'$ & Bank angle correction magnitude [$^\circ$] \\
$\sigma_0$ & Fully numeric predictor-corrector aerocapture guidance initial bank angle [$^\circ$] \\
$\sigma_d$ & Fully numeric predictor-corrector aerocapture guidance phase 2 bank angle [$^\circ$] \\
$\sigma_t$ & Downsampling time spread \\
$\boldsymbol{\sigma}_c$ & Gaussian mixture variational autoencoder cluster variances \\
$\Omega$ & Planet rotation rate [$^\circ$/s] \\
\end{longtable*}}

\section{Introduction}
\lettrine{A}{erocapture} is an alternative to propulsive orbit insertion in which the spacecraft flies through the atmosphere of a planet to capture into orbit. This maneuver is an attractive option for missions to ice giants as it reduces transit time and propellant requirements. As the Planetary Science Decadal Survey~\cite{council_vision_2012} placed high priority on missions to ice giants, the proposed Uranus Orbiter Probe mission investigated aerocapture as an alternative to propulsive orbit insertion \cite{dutta_uranus_2024}. However, the outcome of an aerocapture trajectory is highly sensitive to atmospheric interface state dispersions, especially in flight path angle and velocity \cite{grace_two-stage_2022}. If an improper amount of energy is dissipated during atmospheric flight, the spacecraft can either impact the surface of the planet or escape from the planet's sphere of influence back into interplanetary space. This is not a concern for missions with precise navigation solutions, but less expensive missions may employ less accurate navigation solutions. For sufficiently large initial state errors and environment uncertainty, vehicle parameter dispersions, or large atmosphere uncertainty, some portions of the spacecraft's state distribution could lead to a failed trajectory, even if the predicted trajectory mean successfully captures. Aerocapture can result in two unsuccessful outcomes: (1) escape, where the vehicle does not dissipate enough energy to capture into the planet's gravity well, and exits the atmosphere on a hyperbolic trajectory, and (2) impact, where the vehicle dissipates too much energy and impacts the surface of the planet rather than successfully exiting the atmosphere.

A variety of predictor-corrector guidance algorithms have been developed for aerocapture guidance. These algorithms have been shown to successfully adapt to large trajectory dispersions, without the need for preplanned reference trajectories. Analytical predictor-correctors make use of closed-form approximations of the dynamics to predict trajectory outcome when computing optimal guidance commands~\cite{chen_augmented_2024,cihan_analytical_2021,chadalavada_desensitized_nodate}. Numeric predictor correctors (NPCs), such as the fully numeric predictor-corrector aerocapture guidance algorithm (FNPAG), use numerical integration to predict trajectory outcome for guidance. However, these algorithms only consider the outcome of the nominal or mean state when computing guidance commands \cite{lu_optimal_2015}. Brunner and Lu have shown that a numeric predictor-corrector guidance algorithm outperforms Apollo skip entry guidance (which has similar flight regimes to aerocapture guidance) for conservative initial state dispersions, though both algorithms were evaluated on 3$\sigma$ velocity dispersions of $\mathcal{O}(10)$ m/s and  3$\sigma$ flight path angle dispersions of roughly 0.4$^\circ$ \cite{brunner_comparison_2012}. While these values are typical for missions to date, there is significant navigation cost to obtain initial dispersions this conservative, especially for outer planet missions \cite{restrepo_mission_2024}.

Additionally, if the probability of the spacecraft belonging to one of the unsuccessful modes is sufficiently high, computing guidance algorithm commands based on the mean state can lead to failures. For example, the onboard guidance models could predict a successful capture when computing guidance commands, while in actuality the vehicle could be escaping or impacting due to environmental dispersions. This is especially pertinent for missions with less accurate navigation systems, as the state and environment uncertainty is higher and the terminal mode (impact, capture, or escape) is more uncertain. In these cases, incorporating a probabilistic indicator function into the guidance algorithm can improve robustness. The proposed probabilistic indicator function is lightweight and efficient, and thus it is simple to integrate into FNPAG without increasing computational cost to enhance robustness for both low- and high-uncertainty missions.

Other flight scenarios, such as skip entry guidance and launch vehicle abort guidance, include deterministic guidance triggers based on predicted flight modes. The nominal guidance mode for the Apollo capsule was a skip entry, but if the predicted skip-out velocity was less than desired, the skip-out was omitted in favor of a direct entry \cite{graves_apollo_1972}. Similarly, guidance for the Orion vehicle transitions between phases during skip entry using deterministic predefined thresholds~\cite{bairstow_reentry_2006}. The Space Shuttle had several ascent abort modes in case of main engine failure, including returning to the launch site and transoceanic abort landing \cite{hage_simulation_1993}. Mission Control selected an abort mode based on vehicle velocity, altitude, and time of engine failure. Risk-aware aerocapture guidance extends these concepts to an autonomous, probabilistic framework to increase capture rate for aerocapture missions with high navigation uncertainty.

A variety of techniques have been employed to improve aerocapture guidance in uncertain conditions. Previous work has focused on improving onboard density estimation within the FNPAG framework using recurrent neural networks and the Karhunen-Lo\`eve expansion \cite{rataczak_density_nodate, sonandres_real-time_2025, albert_dimensionality_2025}. Others have reformulated the entry and aerocapture guidance problems to use convex optimization techniques to allow more flexibility in the objective function and constraints \cite{rataczak_convex_nodate, tracy_cpeg_2022, zucchelli_two_nodate}. Optimal control theory using backward reachable sets~\cite{nagabhushana_aerocapture_2025} and guidance based on machine learning~\cite{ghiugan_uranus_nodate} have been applied to aerocapture guidance. While these approaches can improve guidance performance and account for uncertainty, they focus on relatively small initial state dispersions and do not include failure modes (escape or impact) in the guidance decision-making process. 

In the present work, a risk-aware aerocapture guidance algorithm is introduced, which uses a probabilistic indicator of trajectory mode to inform and bias guidance commands. This paper considers low lift-to-drag-ratio vehicles employing bank angle steering in the FNPAG predictor-corrector structure. This algorithm consists of two main components: the probabilistic indicator function and the corrective guidance command. A Gaussian Mixture Variational Autoencoder (GMVAE), which is a computationally efficient clustering algorithm, will be used as the probabilistic indicator function. Once trained, a GMVAE can efficiently and accurately determine modal probabilities for a given trajectory~\cite{dilokthanakul_deep_2017}. When finding the optimal bank angle to minimize target apoapsis error, the probabilistic indicator function is evaluated to determine the probability of each mode. If the modal probability of impact or escape is too high, or the modal probability of capture is too low, the commanded bank angle is biased away from the predicted failure mode. If none of these thresholds are exceeded, the command from the numeric predictor-corrector guidance algorithm is executed.

The remainder of this paper is organized as follows. First, we outline the aerocapture problem and specifics of the scenarios investigated in this paper. Second, the FNPAG implementation, including an improved NPC objective function, are discussed. Third, we detail the training and validation of the GMVAE probabilistic indicator function. Fourth, we present the risk-aware aerocapture guidance algorithm, referred to as $\pi$PAG. Fifth, we present the results of the risk-aware aerocapture guidance algorithm and compare it to the augmented FNPAG algorithm in numerous Monte Carlo experiments. Finally, we close with a discussion of results and concluding remarks.

\section{The Aerocapture Problem}
\label{sec:problem}

The equations of motion over an oblate-spheroid, rotating planet for a spacecraft in the atmosphere are given by the following equations \cite{miele_optimal_1989}:

\begin{equation} \label{eqn:pos}
\frac{dr}{dt} = V \sin\gamma
\end{equation}
\begin{equation}
\frac{d\theta}{dt} = \frac{V \cos\gamma \cos\psi}{r \cos\phi}
\end{equation}
\begin{equation}
\frac{d\phi}{dt} = \frac{V \cos\gamma \sin\psi}{r}
\end{equation}
\begin{equation}
\frac{dV}{dt} = F_T - g_r \sin\gamma - g_\phi \cos\gamma \sin\psi 
+ \Omega^2 r \cos\phi (\sin\gamma \cos\phi - \cos\gamma \sin\phi \sin\psi)
\end{equation}
\begin{equation}
\begin{split}
\frac{d\gamma}{dt} = \frac{1}{V} &\Big(F_N \cos\sigma - g_r \cos\gamma + \frac{V^2}{r} \cos\gamma 
+ g_\phi \sin\gamma \sin\psi\\
&+ 2\omega V \cos\phi \cos\psi 
+ \Omega^2 r \cos\phi (\cos\gamma \cos\phi + \sin\gamma \sin\phi \sin\psi)\Big)
\end{split}
\end{equation}
\begin{equation}
\begin{split}
\frac{d\psi}{dt} = \frac{1}{V} &\Bigg(\frac{F_N \sin\sigma}{\cos\gamma} - \frac{V^2}{r \cos\gamma \cos\psi} \tan\phi 
- \frac{g_\phi \cos\psi}{\cos\gamma} \\
&+ 2\Omega V (\tan\gamma \cos\phi \sin\psi - \sin\phi) 
- \frac{\Omega^2 r \sin\phi \cos\phi \cos\psi}{\cos\gamma}\Bigg)
\end{split}
\end{equation}
\begin{equation} \label{eqn:bank}
\frac{d\sigma}{dt} = \frac{\sigma^* - \sigma}{\zeta},
\end{equation}
where $r$ is the vehicle position radius relative to the center of the Earth, $\theta$ is the longitude, $\phi$ is the latitude, $V$ is the planet-relative velocity, $\gamma$ is the flight path angle (defined positive above local horizon), $\psi$ is the heading angle, $\Omega$ is the rotation rate of the planet, $\sigma$ is the current bank angle, $\sigma^*$ is the commanded bank angle, and $\zeta$ is the bank angle time constant. The radial and latitudinal components of gravity are 
\begin{equation} 
	g_r=\frac{\mu}{r^2}\left(1+J_2\left(\frac{R_e}{r}\right)^2\left(1.5-4.5 \sin ^2 \phi\right)\right),
\end{equation}
\begin{equation}
g_\phi=3 \frac{\mu}{r^2}\left(J_2\left(\frac{R_e}{r}\right)^2 \sin \phi \cos \phi\right),
\end{equation}
where $\mu$ is the gravitational parameter of the planet, $J_2$ is the second zonal harmonic of the planet, and $R_e$ is the equatorial radius of the planet. 

The aerodynamic accelerations are given by:
\begin{equation} \label{eqn:drag}
	F_T = - \frac{\rho V^2}{2\beta},
\end{equation}
\begin{equation}  \label{eqn:lift}
	F_N = \frac{\rho V^2 \frac{L}{D}}{2\beta}, 
\end{equation}
where $\rho$ is the atmospheric density, $\beta = \frac{m}{C_D A_{\text{ref}}}$ is the ballistic coefficient, $\frac{L}{D} = \frac{C_L}{C_D}$ is the lift-to-drag ratio.

In the truth model, the bank angle changes per \cref{eqn:bank}. The bank angle rate is constrained to be less than 20 $^\circ$/s, and if \cref{eqn:bank} calls for a higher bank angle rate, it is clipped to this constraint. 

%The nominal state initial conditions for a Uranus aerocapture are presented in \cref{tab:init_conditions}. These conditions are based on the values used for the Uranus Orbiter Probe mission design \cite{dutta_uranus_2024,deshmukh_6-dof_2025}. Two scenarios are used for this paper\textemdash near-escape and near-impact\textemdash as two Gaussian-distributed datasets are needed to represent all three modes (capture, impact, and escape) in the data. The initial flight path angle of the near-escape trajectory is in the center of the flight path angle corridor, and dispersions result in escaped cases. The initial flight path angle for the near-impact trajectory is at the bottom of the flight path angle corridor such that dispersions result in impacts. Each guidance simulation is run for 1,500 s with a 1 s time-step. The nominal vehicle has a ballistic coefficient $\beta$ of 145 kg/m$^2$, a lift-to-drag ratio $\frac{L}{D}$ of 0.25, and a nominal mass of 2847.068 kg \cite{dutta_uranus_2024,deshmukh_6-dof_2025}. 

Two atmosphere models are used for this study: Uranus Global Reference Atmospheric Model (UranusGRAM) samples \cite{justh_uranus_2021} and a polynomial fit to UranusGRAM \cite{matz_analysis_2024}. The UranusGRAM samples are drawn for altitudes between 0 and 5,000 km by drawing a random seed from $\mathcal{U}(1, 29999)$ and varying the density perturbation scale, \texttt{dp}. A density perturbation scale of 1 corresponds to 3$\sigma$ dispersions, and a density perturbation scale of 2 corresponds to 6$\sigma$ dispersions. Larger density perturbation scales were chosen to produce the most variability in the training data. The atmosphere is assumed to be uniform across the surface of the planet, so the latitude and longitude are held constant when drawing samples. All UranusGRAM profiles are simulated in advance of guidance simulations, and then a single UranusGRAM profile is used for a given simulation. The density values are linearly interpolated from the predefined grid to the vehicle's altitude. For the onboard model, a simplified polynomial fit was found following \cite{matz_analysis_2024}:
\begin{equation}
    \ln \rho = \frac{c_1 + c_3h+c_2h^2 + c_7h^3 + c_9h^4}{1 + c_2h + c_4h^2 + c_6 h^3 + c_8 h^4}. 
\end{equation}

This higher-order polynomial fit was required rather than the commonly-used exponential because using a single exponential atmosphere fit over all the altitudes in the aerocapture flight regime yields a high mean error between the GRAM average and the exponential fit. The polynomial fit coefficients are obtained by performing a nonlinear least squares curve fit between 200 and 2,000 km altitude, and are included in \cref{tab:poly_atmo_coefficients}. 

\setcounter{table}{0}
\begin{table}[htb]
    \centering
    \caption{Polynomial Atmosphere Model Coefficients.}
    \label{tab:poly_atmo_coefficients}
    \begin{tabular}{ccccccccc}
        \hline
        $c_1$ & $c_2$ & $c_3$ & $c_4$ & $c_5$ & $c_6$ & $c_7$ & $c_8$ & $c_9$ \\
        \hline
        -1.01 & 1.18$\times 10^{7}$ & -8.47$\times 10^{7}$ & -36.1 & 181 & 4.10$\times 10^{-5}$ & -4.78$\times 10^{-5}$ & 1.86$\times 10^{-11}$ & -7.03$\times 10^{-10}$ \\
        \hline
    \end{tabular}
\end{table}

\begin{figure}[hbt]
	\centering
	\begin{subfigure}[b]{0.49\textwidth}
         \centering
         \includegraphics[width=\textwidth]{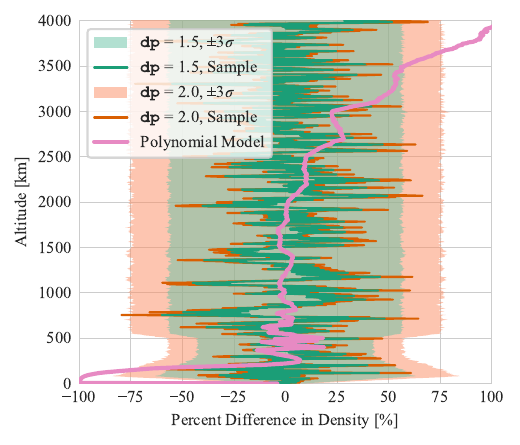}
         \caption{Atmospheric Uncertainty.}
         \label{fig:atmo_uncert}
     \end{subfigure}
     \hfill
     \begin{subfigure}[b]{0.49\textwidth}
         \centering
         \includegraphics[width=\textwidth]{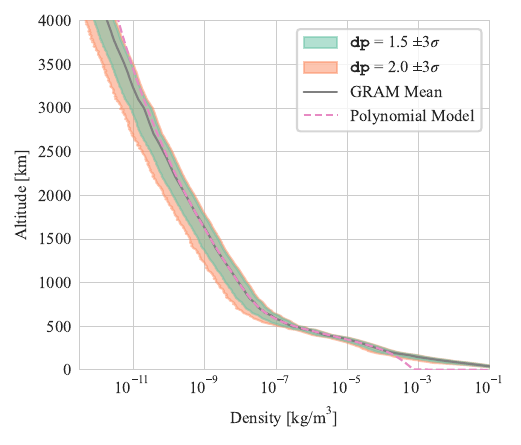}
         \caption{Atmosphere Models.}
         \label{fig:atmo_models}
     \end{subfigure}
     \caption{GRAM and polynominal atmosphere models.}
     \label{fig:atmos}
\end{figure}

Both atmosphere models are compared in \cref{fig:atmos}.

An escaped trajectory is defined as a trajectory with a negative final apoapsis radius, 
\begin{equation}
	r_a = a \left(1 + \sqrt{1 - \frac{V_f^2 r_f^2 \cos^2(\gamma_f)}{\mu a}}\right),
\end{equation}
where $r_f$, $V_f$, and $\gamma_f$ are the radial position, velocity, and flight path angle at atmospheric exit interface, and the semimajor axis $a$ is 
\begin{equation}
	a = \frac{\mu}{\frac{2\mu}{r_f} - V_f^2}.
\end{equation}

An impact trajectory meets one of three conditions: 
\begin{enumerate}
	\item The trajectory has a periapsis radius less than 100 km, where periapsis radius is defined as:
\begin{equation}
	r_p = a \left(1 - \sqrt{1 - \frac{V_f^2 r_f^2 \cos^2(\gamma_f)}{\mu a}}\right),
\end{equation}
	\item The trajectory does not exit the atmosphere, $r_f < r_0$, or
	\item The final radial velocity is inward, $\dot{r}_f < 0$.
\end{enumerate}

A captured trajectory is a trajectory which meets none of these conditions (i.e., a trajectory which exits the atmosphere with a positive apoapsis radius, periapsis altitude above 100 km, and positive altitude rate). 

The required $\Delta v$ to correct an apoapsis and periapsis miss is defined by 
\begin{equation}
\begin{split}
    & \Delta v_1 + \Delta v_2  =  \\ & \sqrt{2\mu} \left( \left| \sqrt{\frac{1}{r_a} - \frac{1}{r_a + r_{p,\text{targ}}}} - \sqrt{\frac{1}{r_a} - \frac{1}{2a}} \right|  + \left| \sqrt{\frac{1}{r_{p,\text{targ}}} - \frac{1}{r_{a,\text{targ}}  +  r_{p,\text{targ}}}} - \sqrt{\frac{1}{r_{p,\text{targ}}} - \frac{1}{r_a + r_{p,\text{targ}}}} \right| \right),
\end{split}
\end{equation}
where $\Delta v_1$ corrects the apoapsis error and $\Delta v_2$ corrects the periapsis error. The required $\Delta v$ to correct an inclination (out-of-plane) error is
\begin{equation}
	\Delta v_i = 2\, v_{a,\text{targ}} \sin\!\left(\frac{\Delta i}{2}\right),
\end{equation}
where 
\begin{equation}
	\Delta i =  \left| \arccos\!\left(\cos\theta_f \cos\phi_f\right) - \arccos\!\left(\cos i_{\text{targ}}\right) \right|,
\end{equation}
and 
\begin{equation}
	v_{a,\text{targ}} = \frac{\sqrt{\mu \, a_{\text{targ}}\left(1 - e_{\text{targ}}^2\right)}}{r_{a,\text{targ}}},
\end{equation}
where
\begin{equation}
	e_{\text{targ}} = \frac{r_{a,\text{targ}} - r_{p,\text{targ}}}{r_{a,\text{targ}} + r_{p,\text{targ}}},
\end{equation}
and
\begin{equation}
	a_{\text{targ}} = \frac{1}{2} \left( r_{a,\text{targ}} +  r_{p,\text{targ}} \right)
\end{equation}

The total required $\Delta v$ to correct in-plane and out-of-plane trajectory errors is given by:
\begin{equation}
	\Delta v_{\text{tot}} = \Delta v_1 + \Delta v_2 + \Delta v_i.
\end{equation}

\section{Fully Numeric Predictor-Corrector Aerocapture Guidance (FNPAG)}
\label{sec:fnpag}

The FNPAG Algorithm is a state-of-the-art numeric predictor-corrector bank angle modulation aerocapture guidance algorithm~\cite{lu_optimal_2015}. In this study, an augmented version of FNPAG with optimal apoapsis targeting (Mode 1) with modified lateral logic targeting orbit inclination~\cite{smith_predictive_nodate} is employed, as presented in Algorithm \ref{alg:fnpag}. To derive FNPAG, Lu et al. prove that the optimal solution to the apoapsis targeting problem for aerocapture has a bang-bang structure, assuming that 1) the objective function is only a function of the terminal longitudinal variables ($r_f, V_f, \gamma_f$) and 2) bank angle is the only control variable \cite{lu_optimal_2015}.\footnote{Lu et al. also assume a non-rotating planet and spherical gravity model.} Because of this, an FNPAG trajectory is divided into two phases. In phase 1, the vehicle flies at a small initial bank angle $\sigma_0$ until the optimal switching time, after which guidance assumes the vehicle flies at a larger bank angle $\sigma_d$ in phase 2. 
%{\color{red} Finally, note that sigma\_d is what is assumed to be flown by FNPAG during phase 1; in phase 2, the bank angle flown is computed online, such that a constant bank angle reaches the desired apoapsis. }

When the truth and guidance models are identical, the bang-bang control structure is sufficient to reach the apoapsis target. However, discrepancies between the truth and the onboard guidance model result in the vehicle failing to achieve the desired apoapsis target by maintaining a constant $\sigma_d$ in phase 2. This occurs because the optimal $\sigma_d$ is found without accounting for the true parameters. To address this, in phase 2, a bank angle that minimizes apoapsis radius error is found at each guidance cycle. This enables FNPAG to adapt to differences between the onboard guidance models and the truth models. To find the optimal switching time in phase 1 and the optimal bank angle in phase 2, a root-finding problem is solved using initial bracketing and Brent's Method \cite{brent_algorithms_2013}.

\begin{algorithm}[htb]
	\caption{Fully Numeric Predictor-Corrector Aerocapture Guidance (FNPAG) Algorithm for Apoapsis Targeting~\cite{lu_optimal_2015}.}

	\label{alg:fnpag}
	\KwIn{Initial state $\mathbf{x}_0$, guidance model \texttt{guidance\_model}, truth model \texttt{truth\_model}}
	
	Initialize $t_\text{switch} \gets 300$ s, \texttt{gLim} $\gets 0.1$ Earth g, $\sigma_0 \gets 10^\circ$, $\sigma_d \gets 90^\circ$\;
	Set $\texttt{guidEnabled} \gets \text{False}$, \texttt{FNPAG\_phase} $\gets 1$\;

	\While{$t_k < t_f$}{
        Update fading filters for density estimation using \cref{eqn:drag_ff,eqn:lift_ff}\;

		\If{$g\geq \texttt{gLim}$}{
			$\texttt{guidEnabled} \gets \text{True}$\;
		}
		\Else{
			$\texttt{guidEnabled} \gets \text{False}$\;
		}
		
		\If{$\texttt{guidEnabled}$}{
			Compute $t_\text{switch}$ (Phase 1) or $\sigma^*$ (Phase 2) by solving root-finding problem using \cref{eqn:energy} with guidance model \texttt{guidance\_model} and past state $\mathbf{x}_{k-1}$\; 
		}

        Compute bank angle direction based on lateral logic targeting inclination\;

		\Comment{Get Bank Angle command based on current phase}
		\If{$t < t_\text{switch}$}{
			$\sigma_{command} \gets \sigma_0$\;
		}
		\Else{
			$\sigma_{command} \gets \sigma^*$\; \label{alg:gmvae_insert}
		}

		\Comment{Integrate truth dynamics in Eqs. \ref{eqn:pos}-\ref{eqn:bank} to update state}
		Update $\mathbf{x}_{k+1} \gets \text{Integrate}(\mathbf{x}_k, \sigma_{command}, \texttt{truth\_model})$\;

		Set $\mathbf{x}_k$ = $\mathbf{x}_{k+1}$, $k = k+1$, $t_k = t_k + \delta t$\;

		\Comment{Check for phase switch}
		\If{$t > t_\text{switch}$ \text{and} $\texttt{FNPAG\_phase} == 1$}{
			\texttt{FNPAG\_phase} $\gets 2$\;
		}
	}
\end{algorithm}

Although many aerocapture studies determine the guidance command by minimizing a function of the apoapsis radius error, in this paper, we adopt an alternative objective function derived by equating the specific energy at atmospheric exit to the specific energy corresponding to the target apoapsis radius, assuming $J_2$ is negligible \cite{rataczak_convex_nodate}:
\begin{equation} \label{eqn:energy}
    0 = \left(\frac{1}{r_f} - \frac{V_f^2}{2}\right) - \left(\frac{1}{r_{a, \text{target}}} - \frac{r_f^2 V_f^2 \cos^2 \gamma_f}{2 r_{a, \text{target}}^2} \right).
\end{equation}
A Monte Carlo analysis comparing Lu et al.'s apoapsis objective function to Rataczak et al.'s energy objective function is presented in Appendix \ref{app:fnpag_obj}. Using this objective function for the high-uncertainty scenarios investigated in this paper results in a higher percentage of captured trajectories. To improve numerical stability, all position quantities are nondimensionalized by the planet's equatorial radius, $R_e$, and velocity quantities are nondimensionalized by $\sqrt{R_e g_0}$ following the convention of Lu et al. \cite{lu_entry_2014}. The nondimensionalized quantities are also used in \cref{eqn:energy}. When using the energy NPC objective function~\cite{rataczak_convex_nodate}, alternative lateral logic~\cite{smith_predictive_nodate}, and nondimensional dynamics, the algorithm is referred to as $\varepsilon$FNPAG.

\begin{figure}[hbt]
	\centering
	\includegraphics[width=\textwidth]{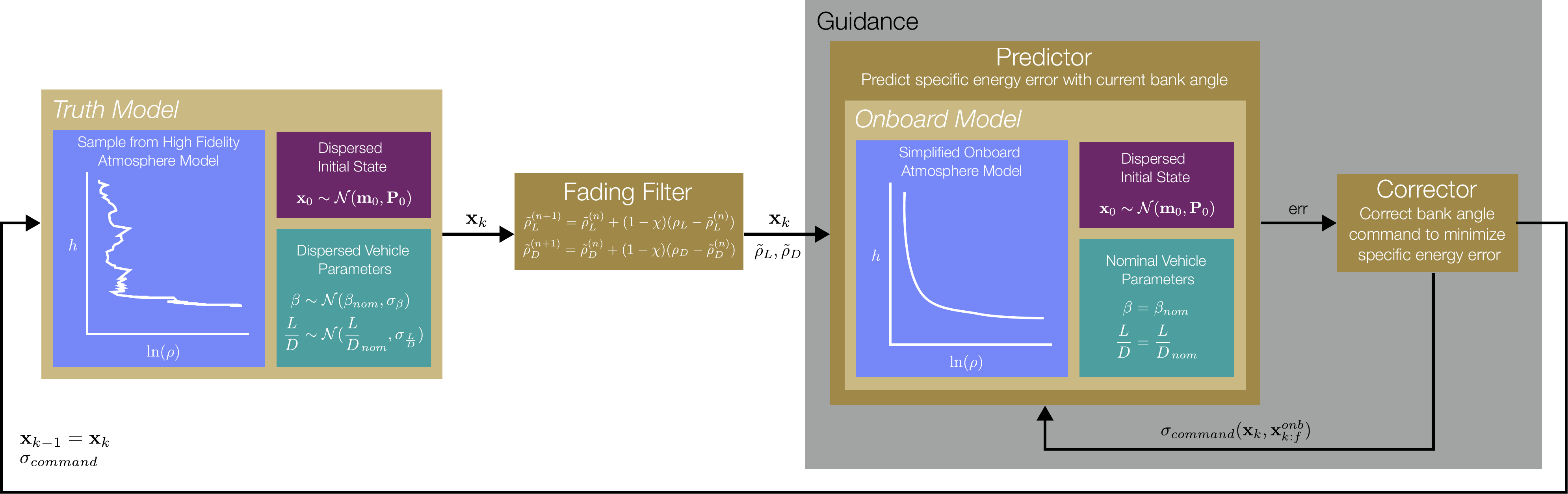}
  \caption{\eFNPAG structure and models used in guidance simulations.}
  \label{fig:fnpag_flow}
\end{figure}

The guidance simulation consists of two models---the truth model and the onboard guidance model---as shown in \cref{fig:fnpag_flow}. The truth model incorporates dispersions in vehicle parameters, atmospheric conditions, and initial conditions. While the guidance model accounts for dispersed initial conditions under the assumption of perfect navigation, it does not incorporate dispersions in vehicle and atmospheric properties. Guidance is not activated until the aerodynamic accelerations exceed a threshold of 0.1 g when the vehicle has enough control authority, and guidance is terminated when the aerodynamic acceleration fall below the 0.1 g threshold. The initial bank angle, $\sigma_0$, is set to 10$^\circ$ and the larger bank angle, $\sigma_d$, is set to 90$^\circ$. Prior studies show FNPAG performance is largely insensitive to these parameters, so values with strong performance across a range of planets and initial conditions were selected~\cite{rataczak_convex_nodate}. The guidance algorithm is run until either the final time, $t_f=1500$ s, is reached or the vehicle's altitude is negative. This final time ensures that all captured and escaped trajectories have exited the atmosphere.

To improve onboard density estimation onboard, two first-order fading-memory filters using IMU data are implemented~\cite{lu_optimal_2015,brunner_skip_2008}. The filters estimate the ratios:
\begin{align}
    \rho_L &= \frac{L}{L^*}, \\ 
    \rho_D &= \frac{D}{D^*}, 
\end{align}
where $L$ and $D$ are the true lift and drag forces measured by the IMU and $L^*$ and $D^*$ are the lift and drag forces computed using the onboard model. In the $(n+1)$-th guidance cycle, the navigation scaling factors, $\tilde{\rho}_L^{(n+1)}$ and $\tilde{\rho}_D^{(n+1)}$, are updated following:
\begin{align} \label{eqn:drag_ff}
    \tilde{\rho}_L^{(n+1)} &= \tilde{\rho}_L^{(n)} + (1 - \chi) (\rho_L - \tilde{\rho}_L^{(n)}), \qquad 0 < \chi < 1 \\ \label{eqn:lift_ff}
    \tilde{\rho}_D^{(n+1)} &= \tilde{\rho}_D^{(n)} + (1 - \chi) (\rho_D - \tilde{\rho}_D^{(n)}), \qquad 0 < \chi < 1
\end{align}
where $\rho_L$ and $\rho_D$ are based on the current IMU measurement, $\tilde{\rho}_L^{(n)}$ and $\tilde{\rho}_D^{(n)}$ are the previous estimates, and $\chi$ is the filter time constant. The initial values for $\tilde{\rho}_L^{(0)}$ and $\tilde{\rho}_D^{(0)}$ are set to 1, and $\chi = \exp(-1/6) \approx 0.84$ is used. A value of $\chi$ near 1 is chosen to emphasize past estimates over the most recent estimates~\cite{brunner_skip_2008}. The resultant $\tilde{\rho}_L^{(n+1)}$ and $\tilde{\rho}_D^{(n+1)}$ are used to scale the lift and drag accelerations in \cref{eqn:drag,eqn:lift} in onboard numerical integrations during the prediction step.

\section{Gaussian Mixture Variational Autoencoder Training and Evaluation}

This section introduces the concept of a probabilistic indicator function and the GMVAE neural network structure, training data, and performance evaluation. We define a probabilistic indicator function as a function that determines class membership probabilistically. While a traditional indicator function maps to one for elements which belong to a given set and zero for members not belonging to that set, a probabilistic indicator function determines the probability of membership to each set for a given sample. 

The choice of a reliable probabilistic indicator function for guidance depends on function accuracy, evaluation efficiency, and ease of implementation in guidance. Thus, a good probabilistic indicator function needs to use the information onboard the spacecraft to determine trajectory outcome probability efficiently and accurately. The available information for decision-making onboard the spacecraft are the previous points along the trajectory (the ``as-flown'' trajectory) and our prediction of the trajectory outcome using onboard models (the ``predicted'' trajectory). Constructing a probability density function (pdf) of guided aerocapture trajectory outcome given this trajectory information is nontrivial. Thus, we selected a data-driven unsupervised clustering algorithm to construct an outcome pdf given the current as-flown and predicted trajectory. While numerous clustering algorithms exist~\cite{xu_comprehensive_2015}, we chose GMVAEs for their ability to capture nonlinear relationships in the data and efficiently estimate cluster membership probabilities. GMVAEs have been previously applied to pattern recognition problems~\cite{dilokthanakul_deep_2017}, peptide classification~\cite{varolgunes_interpretable_2020}, and flow characterization~\cite{fan_physically_2025}. 

\cref{sec:gmvae_arch} introduces the GMVAE training objective function, \cref{sec:training_data} describes training data generation, and \cref{sec:hyperparameter} presents the results of a hyperparameter sweep and the criteria for the selection of a GMVAE to use as a probabilistic indicator function.

\subsection{GMVAE Architecture}
\label{sec:gmvae_arch}
A GMVAE is an unsupervised clustering and dimensionality reduction algorithm that uses a Gaussian Mixture Model (GMM) to cluster data in a lower-dimensional latent space. GMVAEs are an extension of Variational Autoencoders (VAEs). A VAE is a neural network architecture based on variational Bayesian methods that encodes a large dimensional input, $\mathbf{x}\in\mathbb{R}^n$, (i.e. the aerocapture trajectory) into a lower dimensional latent space, $\mathbf{z}\in\mathbb{R}^d$, where $d \ll n$, according to a given distribution, and then decodes the information in the latent space back to the input space~\cite{cinelli_variational_2021}. The use of deep neural networks for the encoder and decoder allows the autoencoder to leverage nonlinear relationships in the input data. 

\begin{figure}[htb]
    \centering
    \includegraphics{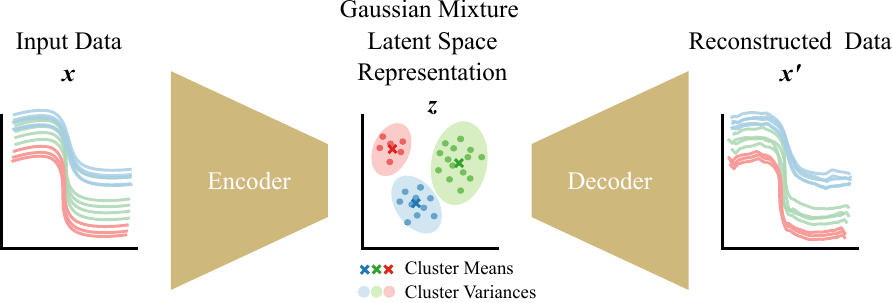}
    \caption{GMVAE schematic for two-dimensional latent space with three data clusters.}
    \label{fig:gmvae_structure}
\end{figure}

A GMVAE differs from a traditional VAE as the latent space is a mixture of Gaussians instead of a single Gaussian, which allows the GMVAE to capture multi-modal distributions in the input data, depicted notionally in \cref{fig:gmvae_structure}. After encoding input data to the latent space, the GMVAE can determine the probability that a given sample belongs to a certain cluster using the weights, means, and covariances of the GMM components. By reducing the dimension of the data from $n$ to $d \ll n$, the GMVAE can perform more efficient clustering on the lower dimensional encoded data. Only the encoded latent space representation is required to determine the cluster membership probabilities, while the decoder is required for training to ensure that the latent space is an accurate representation of the input data.

The main advantage of using a GMVAE is the computational speed with which trajectory cluster membership can be determined. The GMVAE is trained to minimize the log-evidence lower boundary (ELBO) \cite{dilokthanakul_deep_2017}:
\begin{equation} \label{eqn:ELBO}
    \text{ELBO} = \mathbb{E}_q\left[\frac{p_{\lambda,\kappa}(\mathbf{x}',\mathbf{x},c,\mathbf{z})}{q(\mathbf{x},c,\mathbf{z}|\mathbf{x}')} \right],
\end{equation} 
where $\mathbf{x}'$ is the reconstructed data, $\lambda$ and $\kappa$ are the decoder and encoder parameters, $p_{\lambda,\kappa}$ is the joint distribution of the data and latent space, and $q$ is the variational posterior. The ELBO measures how well the latent space is captured by the GMM and how well the decoder reconstructs the input data. This work employs a novel GMVAE framework, which estimates cluster probability through an Expectation Maximization (EM) algorithm as part of the training \cite{fan_physically_2025}. In this formulation, the GMVAE learns the weights, means, and covariances of the latent space GMM representation of the data in latent space. In the expectation step, the cluster membership probabilities for sample $i$, $\gamma_{ic}$ are found by computing the probability of a certain latent variable given a certain cluster, $p(\mathbf{z}_i|c)$. In the maximization step, the GMM component means and variances of cluster $c$, $\boldsymbol{\mu}_c$ and $\boldsymbol{\sigma}_c$, are computed based on each $\gamma_{ic}$. A forward evaluation of the trained encoder neural network and evaluation of the expectation step is all that is needed to determine the modal probabilities for a sample $\mathbf{x}$. 

To construct the ELBO expression in \cref{eqn:ELBO}, the latent prior $p(\mathbf{z})$ is modeled as a GMM
\begin{equation} \label{eqn:gmvae_pz}
    p(\mathbf{z}) = \sum_c \pi_c \mathcal{N}(\mathbf{z}|\boldsymbol{\mu}_c, \boldsymbol{\sigma}^2_c),
\end{equation}
where $\pi_c$ is the cluster weight or probability (learned via gradient descent during training~\cite{fan_physically_2025}), $\boldsymbol{\mu}_c$ is the cluster mean, and $\boldsymbol{\sigma}^2_c$ is the cluster variance (both learned during the EM step). 

Using the GMM structure of the latent space and simplifying the ELBO expression, \cref{eqn:ELBO} can be written as~\cite{dilokthanakul_deep_2017}:
\begin{equation} \label{eqn:simplified_elbo}
    \begin{split}
    \text{ELBO} &= \mathbb{E}_{q(\mathbf{x}|\mathbf{x}')}[\log{p_\kappa(\mathbf{x}'|\mathbf{x})}] - \mathbb{E}_{q(c|\mathbf{x}')p(\mathbf{z}|\mathbf{x},c)}[\text{KL}\left(q_{\phi_x}(\mathbf{x}|\mathbf{x}')||p_\lambda(\mathbf{x}|c,\mathbf{z})\right)] \\
    & - \text{KL}(q_{\phi_c}(c|\mathbf{x}')||p(c)) - \mathbb{E}_{q(\mathbf{x}|\mathbf{x}')q(c|\mathbf{x}')}\left[\text{KL}\left(p_\lambda(\mathbf{z}|\mathbf{x},c)||p(\mathbf{z}) \right) \right],
    \end{split}
\end{equation}
where $\phi_x$ and $\phi_c$ are recognition networks that parametrize each variational factor, and $\text{KL}(\cdot||\cdot)$ denotes the Kullback-Leibler (KL) divergence. The ELBO expression in \cref{eqn:simplified_elbo} is then optimized using the following EM algorithm. To find the cluster membership probabilities for sample $\mathbf{x}$, posterior probability $p(c|\mathbf{z})$ is evaluated via Bayes rule:
\begin{equation}
    p(c|\mathbf{z}) = \frac{p(\mathbf{z}|c)p(c)}{p(\mathbf{z})},
\end{equation}
where $\mathbf{z}$ is the sample of latent variables corresponding to $\mathbf{x}$, and the prior probability of cluster $c$ is $p(c) = \pi_c$. The likelihood of the latent sample $\mathbf{z}$ under cluster $c$ is given by $p(\mathbf{z}|c) = \mathcal{N}(\mathbf{z}|\boldsymbol{\mu}_c, \boldsymbol{\sigma}^2_c)$. The total likelihood of $\mathbf{z}$ given the current parameters is $p(\mathbf{z}) = \sum_k \pi_k \mathcal{N}(\mathbf{z}|\boldsymbol{\mu}_k, \boldsymbol{\sigma}^2_k)$.

%The terms in this expression are referred to as the reconstruction loss, conditional prior term, cluster weight prior term, and latent variable prior term respectively. 

The membership probabilities for a sample $\mathbf{x}_i$ are computed in the expectation step as follows:
\begin{enumerate}
    \item {Encode} the sample $\mathbf{x}_i$ to get the mean $\boldsymbol{\mu}_i$ and log variance $\log \boldsymbol{\sigma}_i^2$ of the latent distribution $q(\mathbf{z}_i|\mathbf{x}_i)$, the approximate posterior over $\mathbf{z}_i$.
    \item {Compute the log probability of the sample belonging to each cluster}: 
    \begin{equation} \label{eqn:log_pz_c}
        \log p(\mathbf{z}_i,c) = \log \pi_c + \log \mathcal{N}(\mathbf{z}_i|\boldsymbol{\mu}_c, \boldsymbol{\sigma}^2_c).
    \end{equation}
    \item {Apply Bayes rule to obtain the posterior $p(c|\mathbf{z}_i)$}:
    \begin{equation} \label{eqn:gamma_c}
        \gamma_{ic} = p(c|\mathbf{z}_i) = \frac{p(\mathbf{z}_i,c)}{\sum_k p(\mathbf{z}_i,k)}.
    \end{equation}
\end{enumerate}

After computing the cluster membership probabilities for each sample $\mathbf{x}_i$, $\gamma_{ic}$, the maximization step updates the global mixture parameters \(\{\boldsymbol{\mu}_c, \boldsymbol{\sigma}_c^2\}\) by maximizing the ELBO with respect to these parameters while holding the cluster probabilities fixed. This is equivalent to maximizing the expected complete-data log-likelihood:
\begin{equation}
\mathcal{L}_{\text{M-step}} = \sum_{i} \sum_{k} \gamma_{ik}\left[\log \pi_k+ \log \mathcal{N}(\mathbf{z}_i \mid \boldsymbol{\mu}_k, \boldsymbol{\sigma}_k^2)\right],
\end{equation}
where $i$ is the latent sample index and $k$ is the cluster index. The EM step is beneficial not only during training as cluster probabilities need not be specified a priori, but also during guidance as it facilitates efficient computation of cluster membership probabilities for sample $\mathbf{x}_i$.

\subsection{Training Data Generation}
\label{sec:training_data}
Onboard decision-making can utilize both the as-flown trajectory, representing true vehicle states under the truth models, and the predicted trajectory generated from onboard models. The as-flown data are the true states the vehicle has passed through, influenced by the truth models, while the predicted data are based on the onboard models given our current state. To probabilistically predict trajectory outcomes from this information, the GMVAE is trained on truth trajectories from closed-loop \eFNPAG simulations, which implicitly encode model discrepancies. Because the training data capture the effects of the truth model, the resulting framework allows the guidance algorithm to account for discrepancies between the truth and onboard models. In contrast, onboard deterministic trajectory prediction provides only a single estimated outcome and does not capture these model differences.

The truth model used to generate training data is intentionally less variable than the truth model used in the simulations presented in the results section (see \cref{sec:results}). The enables the onboard probabilistic indicator to account for a range of discrepancies between the truth and onboard models without requiring execution of a high-fidelity truth model onboard. Because the true environment is not known a priori, training is performed using a simplified truth model, while testing is conducted using a higher-fidelity approximation. This approach enables evaluation of GMVAE performance under previously unseen model variations.

When constructing GMVAE training data, it is beneficial to downsample the data to reduce its dimension as the curse of dimensionality translates to longer neural network training times for larger dimension input data \cite{koppen2000curse}. For example, a 1500 s \eFNPAG simulation with a 1 s time step produces a trajectory in $\mathbb{R}^{1500 \times 7}$. To further reduce the input dimension without losing classification information, the training data uses only the spacecraft's energy:
\begin{equation}
    \varepsilon(t) = \frac{1}{2} V(t)^2 - \frac{\mu}{r(t)}.
\end{equation}
Terminal energy is positive for an escaped trajectory and negative for a captured or impacted one, with impacting trajectories exhibiting a large decrease in terminal energy. While reducing the trajectory input from the full state to this single parameter loses some information, the variation in specific energy is sufficient to classify trajectory mode without necessitating a more complex probabilistic indicator function architecture. The energy data are downsampled to 36 nonuniform time steps and divided by its initial mean to ensure all samples are approximately $\mathcal{O}(1)$. For trajectories that impact before the termination time, the final energy value is used to pad the remaining time indices. 

\begin{figure}[hbt]
	\centering
	\begin{subfigure}[t]{0.49\textwidth}
         \centering
         \includegraphics[width=\textwidth]{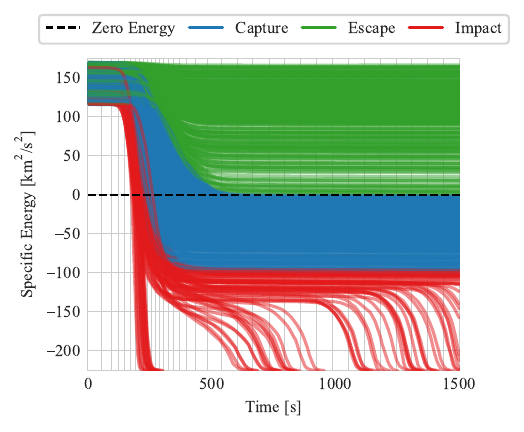}
         \caption{Specific energy of GMVAE training data. X-axis gridlines indicate downsample times.}
         \label{fig:raw_energy}
     \end{subfigure}
     \hfill
     \begin{subfigure}[t]{0.49\textwidth}
         \centering
         \includegraphics[width=\textwidth]{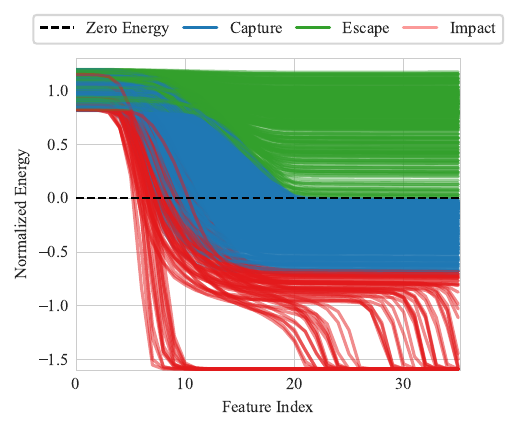}
         \caption{Non-uniformly downsampled and normalized energy training data.}
         \label{fig:training}
     \end{subfigure}
     \caption{GMVAE Training Data.}
     \label{fig:gmvae_training}
\end{figure}

The downsampling grid is chosen such that more samples are taken near the energy gradient during the aerocapture maneuver and fewer samples are taken at the beginning and end of the trajectory. We use inverse transform sampling on a pseudo-Gaussian weighting function to determine the downsampling times. For each $t_i \in \left[ t_0,  t_1,  \ldots,  t_F\right]$, where $t_0 = 0$ s and $t_f = 1500$ s, we define a weight
\begin{equation}
	w_i = 1 + A \exp\!\left(-\frac{(t_i - t_c)^2}{2\sigma_t^2} \right),
\end{equation}
based on the desired central time, $t_c = 250$ s, time standard deviation, $\sigma_t = 100$ s, and cluster weight strength, $A=3$. The desired central time is chosen to be near the energy gradient during the aerocapture maneuver. This function produces a baseline weight of 1 for all time with a psuedo-Gaussian increase in weight around the central time. The discrete cumulative distribution function (CDF) of this distribution is computed as:
\begin{equation}
	F(k) = \frac{\sum_{i=0}^{k} w_i}{\sum_{i=0}^{t_F-1} w_i},
\end{equation}
and 36 uniformly spaced points on the interval $[0,1]$, $u_j$ for $j = 1, \ldots, 36$, are inverted through the CDF to obtain the desired non-uniform in time downsampling grid. The downsampled time indices are obtained by inverting the CDF:
\begin{equation}
	\text{idx}_j = F^{-1}(u_j)
\end{equation}
The resulting index set is more densely concentrated near $t_c$ and more sparsely distributed far from it, producing a non-uniform temporal downsampling biased toward the specified center. The downsampling times resulting from this method are shown as the $x$-axis gridlines in \cref{fig:raw_energy}.

As discussed above, the GMVAE training dataset is constructed to exhibit greater variability than the nominal flight regime tested in the results (\cref{sec:results}) to ensure robust outlier classification. Additionally, the models used to generate this data are chosen to be less extreme than the anticipated variations, allowing evaluation of extrapolation performance. Furthermore, the dataset is enriched with failure cases (escapes and impacts) to improve classification performance in these regimes. To achieve these properties, a Gaussian-Uniform mixture (GU mixture) is used for each dispersed initial state and parameter. The univariate GU mixture for random variable $x$ is defined as:
\begin{equation}
x \sim
\begin{cases}
\mathcal{U}\!\left(\mu - k_{2,\text{low}}\sigma,\; \mu - k_{1,\text{low}}\sigma\right),
& \text{with probability } \dfrac{p}{2}, \\[8pt]

\mathcal{N}\!\left(\mu,\sigma\right),
& \text{with probability } 1 - p, \\[8pt]

\mathcal{U}\!\left(\mu + k_{1,\text{high}}\sigma,\; \mu + k_{2,\text{high}}\sigma\right),
& \text{with probability } \dfrac{p}{2},
\end{cases}
\end{equation}
where $\mu$ is the center point of the distribution, $\sigma$ is the standard deviation, $0 < p < 1$ is the total tail probability, $0 < k_{1,\text{low}} < k_{2,\text{low}}$ define the inner and outer tail bounds for the lower tail in units of $\sigma$, and $0 < k_{1,\text{high}} < k_{2,\text{high}}$ define the inner and outer tail bounds for the upper tail in units of $\sigma$. We selected a GU mixture rather than a single Gaussian or Uniform distribution for the training data generation to control the number of outliers and spread of the training data.

The center points and standard deviations for the initial state and vehicle parameters used to generate the GU mixture samples is presented in \cref{tab:gmave_training_mean}. Note that velocity and flight path angle are given in the inertial frame. The initial flight path angle is chosen to be the center of the flight path angle corridor. The target orbit parameters, vehicle properties, and atmosphere model used in guidance for training are shown in \cref{tab:gmave_training_params}. As described in \cref{sec:problem}, the truth atmosphere model are samples taken from GRAM and the onboard atmosphere model is a polynomial fit to the GRAM mean. 

The spread for most parameters set to be between 2--4 $\sigma$; that is,  $k_{1,\text{low}} = k_{1,\text{high}} = 2$ and $k_{2,\text{high}} = k_{2,\text{low}} = 4$. The negative variance spread is increased for flight path angle, $ k_{2,\text{low}} = 8$, to encourage more impact cases. Including more impact cases in the training data bolsters GMVAE classification accuracy for these cases.  

\begin{table}[htb]
    \centering 
    \caption{Initial center points and standard deviations for GMVAE training data, based on the Uranus Orbiter Probe mission~\cite{deshmukh_6-dof_2025,deshmukh_performance_2024}.}
    \label{tab:gmave_training_mean}
    \begin{tabular}{@{}rcccccccc@{}}
    \hline
       Parameter & $h_0$ [km] & $\theta_0$ [$\deg$] & $\phi_0$ [$\deg$] & $V_{0,I}$ [km/s] & $\gamma_{0,I}$ [$\deg$] & $\psi_0$ [$\deg$] &  $\frac{L}{D}$ [n.d.] & $m$ [kg] \\ \hline
      Center Point & 1000 & 190.045 & -9.764 & 24.936 & -10.572 & 45.00 & 0.25 & 2847.068 \\ 
      $3\sigma$ & 100 & 0.227 & 0.116 & 0.750 & 0.50 & 0.063 & 0.075 & 854.120 \\
      \hline
    \end{tabular}
\end{table}

\begin{table}[htb]
    \centering 
    \caption{Scenario parameters for GMVAE training data.}
    \label{tab:gmave_training_params}
    \begin{tabular}{@{}ccccc@{}}
    \hline
       $h_{a,\text{target}}$ [km] & $h_{p,\text{target}}$ [km] & $i_{\text{target}}$ [$^\circ$] & $\beta$ [kg/m$^2$] & GRAM \texttt{dp} [n.d.] \\ \hline
      550,000 & 4,000 & 45.824 & 145 & 1.5 \\ \hline
    \end{tabular}
\end{table}

1,000 Monte Carlo samples were generated for GMVAE training from initial conditions and atmosphere models sampled as described above. The energy versus time for the training data is shown in \cref{fig:raw_energy}, and the downsampled and normalized energy GMVAE training data is shown in \cref{fig:training}.

%{\color{red} 
%Training data is characterized as having …
%More variability in input conditions
%And Lower variability in atmospheric density
%This Combination creates data that is both more challenging in terms of state variability but less challenging in terms of atmospheric variability.
%}

\subsection{GMVAE Hyperparameter Selection}
\label{sec:hyperparameter}

Tuning GMVAE hyperparameters for accurate trajectory mode classification is nontrivial. The GMVAE must be expressive enough to capture the variability in the input data while avoiding overfitting. Three main hyperparameters must be selected: the latent dimension, $d$, the neural network architecture, and the number of clusters used to fit the latent space, $C$. First, the latent dimension chosen must be large enough to describe the data modes. Second, the selected neural network architecture (i.e. number of hidden layers, hidden layer dimensions, and activation functions) must be sufficiently expressive to capture the variability in the data. Third, the number of clusters must be sufficient to capture non-Gaussian structure in the latent space. While theoretically there are only three clusters to describe the training dataset (capture, escape, and impact), allowing the GMVAE to include more clusters can increase flexibility during training and improve classification accuracy. Each of these tuning parameters is discussed in the following sections.

%{\color{red}
%\begin{itemize}
%	\item SVD for latent dimension 
%	\item Define performance metrics 
%	\item Show results of sweeps in terms of weighted and failure only weighted latent dimensions 
%	\item Discuss that the best GMVAEs were trained for 30,000 epochs on a larger training dataset. 
%\end{itemize}}

\subsubsection{Latent Dimension Selection}

\begin{figure}[hbt]
	\centering
	\begin{subfigure}[b]{0.49\textwidth}
         \centering
		 \includegraphics[width=\textwidth]{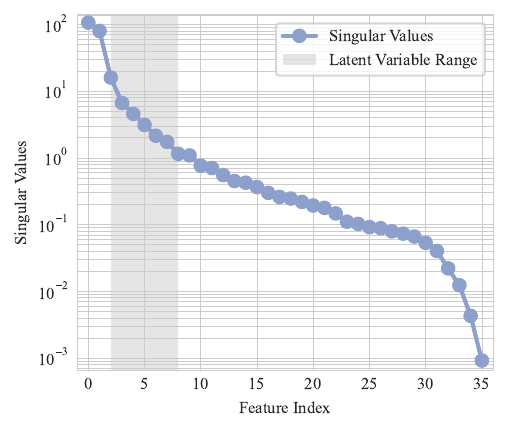}
         \caption{Singular values.}
         \label{fig:singular_values}
     \end{subfigure}
     \hfill
     \begin{subfigure}[b]{0.49\textwidth}
         \centering
         \includegraphics[width=\textwidth]{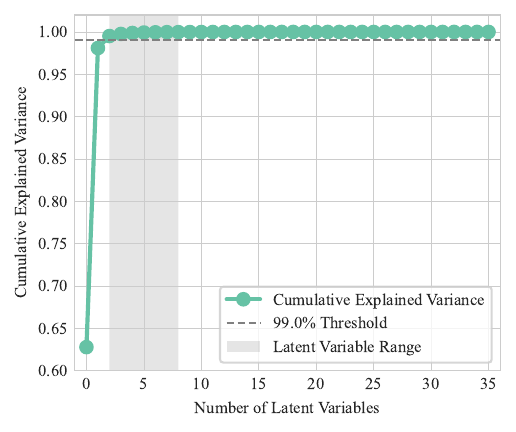}
         \caption{Cumulative explained variance.}
         \label{fig:cev}
     \end{subfigure}
     \caption{Results of SVD of non-uniformly downsampled and scaled energy data.}
     \label{fig:svd}
\end{figure}

To find a relatively small number of latent variables to describe the input data, singular value decomposition (SVD) is performed on the downsampled and scaled energy data. While the autoencoder identifies relevant features in the data, the dimension of the latent space must be specified a priori. By looking at the rate of decrease of the singular value magnitudes, the dimension required to describe the input data without significant loss of information can be determined~\cite{wall_singular_2003}. We perform SVD on the downsampled and scaled energy data, and the singular values are shown in \cref{fig:svd}. The cumulative explained variance, the total proportion of variance captured by the first $k$ singular values, is defined as $\sum_{i=1}^{k} s_i \big/ \sum_{i=1}^{n} s_i$, where $s_i$ are the singular values. This is shown in \cref{fig:cev}, indicating that a latent dimension of at least three is reasonable to describe the input data. We vary the latent dimension from three to nine in the following studies to allow for increased flexibility in training.

\subsubsection{Neural Network Architecture and Cluster Number Selection}
\label{sec:gmvae_training_nn}

While a priori there are three clusters in the dataset, varying the number of clusters in the GMM latent space model can allow the GMVAE more flexibility to fit to the data. The three-layer encoder and decoder architectures evaluated for the GMVAE are shown in \cref{fig:encoder_decoder}. A Gaussian Error Linear Unit (GELU) activation function is used between hidden layers. Network sizes of $d^h_1 \times d^h_2 \times d^h_3$ = $24 \times 18 \times 12$, $30 \times 20 \times 10$, and $36 \times 24 \times 12$ were evaluated. 

% Define colors (RGB)
\definecolor{hiddenblue}{RGB}{113,137,255}
\definecolor{latentgreen}{RGB}{32,160,160}

\begin{figure}[htb]
\centering

% ---------------- Encoder ----------------
\begin{subfigure}[b]{\textwidth}
\centering
\begin{tikzpicture}[
  arrow/.style={->, thick},
  labelstyle/.style={midway, below, font=\small\itshape}
]

% Encoder layers with explicit heights
\node[rectangle, draw, rounded corners,
      fill=layerfill,
      minimum width=2.8cm,
      minimum height=2.0cm,
      align=center] (enc_input)
      {Input (Energy)\\$[36]$};

\node[rectangle, draw, rounded corners,
      fill=hiddenblue!75,
      minimum width=2.5cm,
      minimum height=1.8cm,
      align=center,
      right=1cm of enc_input] (enc_hidden1)
      {Hidden\\$[d^h_1]$};

\node[rectangle, draw, rounded corners,
      fill=hiddenblue!75,
      minimum width=2.3cm,
      minimum height=1.6cm,
      align=center,
      right=1cm of enc_hidden1] (enc_hidden2)
      {Hidden\\$[d^h_2]$};

\node[rectangle, draw, rounded corners,
      fill=hiddenblue!75,
      minimum width=2.1cm,
      minimum height=1.4cm,
      align=center,
      right=1cm of enc_hidden2] (enc_hidden3)
      {Hidden\\$[d^h_3]$};

\node[rectangle, draw, rounded corners,
      fill=latentgreen!85,
      minimum width=1.9cm,
      minimum height=1.0cm,
      align=center,
      right=1cm of enc_hidden3] (enc_latent)
      {Latent\\$\mathbf{z} \in \mathbb{R}^d$};

\draw[arrow] (enc_input) -- node[labelstyle] {GELU} (enc_hidden1);
\draw[arrow] (enc_hidden1) -- node[labelstyle] {GELU} (enc_hidden2);
\draw[arrow] (enc_hidden2) -- node[labelstyle] {GELU} (enc_hidden3);
\draw[arrow] (enc_hidden3) -- node[labelstyle] {GELU} (enc_latent);

\end{tikzpicture}
\caption{Encoder architecture.}
\end{subfigure}

\vspace{1em}
% ---------------- Decoder ----------------
\begin{subfigure}[b]{\textwidth}
\centering
\begin{tikzpicture}[
  arrow/.style={->, thick},
  labelstyle/.style={midway, below, font=\small\itshape}
]

\node[rectangle, draw, rounded corners,
      fill=latentgreen!85,
      minimum width=1.9cm,
      minimum height=1.0cm,
      align=center] (dec_latent)
      {Latent\\$\mathbf{z} \in \mathbb{R}^d$};

\node[rectangle, draw, rounded corners,
      fill=hiddenblue!75,
      minimum width=2.1cm,
      minimum height=1.4cm,
      align=center,
      right=1cm of dec_latent] (dec_hidden1)
      {Hidden\\$[d^h_3]$};

\node[rectangle, draw, rounded corners,
      fill=hiddenblue!75,
      minimum width=2.3cm,
      minimum height=1.6cm,
      align=center,
      right=1cm of dec_hidden1] (dec_hidden2)
      {Hidden\\$[d^h_2]$};

\node[rectangle, draw, rounded corners,
      fill=hiddenblue!75,
      minimum width=2.5cm,
      minimum height=1.8cm,
      align=center,
      right=1cm of dec_hidden2] (dec_hidden3)
      {Hidden\\$[d^h_1]$};

\node[rectangle, draw, rounded corners,
      fill=layerfill,
      minimum width=2.8cm,
      minimum height=2.0cm,
      align=center,
      right=1cm of dec_hidden3] (dec_output)
      {Output (Energy)\\$[36]$};

\draw[arrow] (dec_latent) -- node[labelstyle] {GELU} (dec_hidden1);
\draw[arrow] (dec_hidden1) -- node[labelstyle] {GELU} (dec_hidden2);
\draw[arrow] (dec_hidden2) -- node[labelstyle] {GELU} (dec_hidden3);
\draw[arrow] (dec_hidden3) -- node[labelstyle] {GELU} (dec_output);

\end{tikzpicture}
\caption{Decoder architecture.}
\end{subfigure}

\caption{GMVAE encoder and decoder architectures with two hidden layers.}
\label{fig:encoder_decoder}
\end{figure}

To find the best combination of hyperparameters for the GMVAE, the latent dimensions and number of clusters were varied for each architecture. Both the latent dimension $d$ and number of clusters $C$ were varied between three and nine. To compare each parameter combinations, neural network training was set to be deterministic. Each of the 147 GMVAEs was trained for 10,000 epochs with a batch size of 128 and a learning rate of 0.001. The 1,000-sample Monte Carlo training data were split into 800 training samples, 100 validation samples, and 100 test samples. The Adam optimizer was used to optimize the GMVAE loss function, defined as the ELBO in \cref{eqn:ELBO}. To ensure each cluster membership probability remained above zero, a regularization term of 1$\times 10^{-6}$ was added to $\gamma_{ic}$ in \cref{eqn:gamma_c} after each EM step. 

After training, to determine which mixands belong to which true data cluster, the Mahalanobis distance from every encoded sample to each cluster is computed for all samples. The mixand is assigned to the cluster with the minimum mean Mahalanobis distance across all samples. 

To be effective as a probabilistic indicator function in $\pi$PAG, the GMVAE must accurately label the data. In particular, it is important that the GMVAE can accurately recognize failures (escape and impact). Thus, the weighted misassignment rate and failure-only weighted misassignment rates were used to evaluate GMVAE performance. The GMVAE misassignment rate for a given outcome is computed as:\begin{equation}
e_{\text{outcome}}(d, C) = \frac{1}{N_{\text{outcome}}} \sum_{i=1}^{N} \mathbf{1}\!\left( l_i = {\text{outcome}} \right)\,\mathbf{1}\!\left( \hat{l}_i(d, C) \neq {\text{outcome}} \right)
\end{equation}
where $N_{\text{outcome}}$ is the number of samples with that outcome, $l_i$ is the true outcome label of sample $i$, $\hat{}l_i(d,C)$ is the GMVAE-predicted cluster label for latent dimension $d$ and number of clusters $C$, and $\mathbf{1}(\cdot)$ denotes the indicator function.

The weighted misassignment rate is defined as for each latent dimension $d$ and number of clusters $C$ as: 
\begin{equation} \label{eqn:weighted_rate}
\bar{e}_{\text{W}}(d, C) = P_{\text{cap}} \, e_{\text{cap}}(d, C) + P_{\text{esc}} \, e_{\text{esc}}(d, C) + P_{\text{imp}} \, e_{\text{imp}}(d, C),
\end{equation}
where the cluster probabilities are defined as,
\begin{equation}
P_{\text{cap}} = \frac{N_{\text{cap}}}{N}, 
\quad
P_{\text{esc}} = \frac{N_{\text{esc}}}{N}, \quad \text{and}
\quad
P_{\text{imp}} = \frac{N_{\text{imp}}}{N}, 
\end{equation}
where $N$ is the total number of samples. 

The failure-only weighted misassignment rate only considers mispredictions of the escape and impact modes: 
\begin{equation} \label{eqn:failure_weight}
\bar{e}_{\text{F}}(d,C) = \frac{P_{\text{esc}} \, e_{\text{esc}}(d,C) + P_{\text{imp}} \, e_{\text{imp}}(d,C)}{P_{\text{esc}} + P_{\text{imp}}}.
\end{equation}

\begin{figure}[hbt]
	\centering
	\begin{subfigure}[b]{\textwidth}
         \centering
         \includegraphics[width=\textwidth]{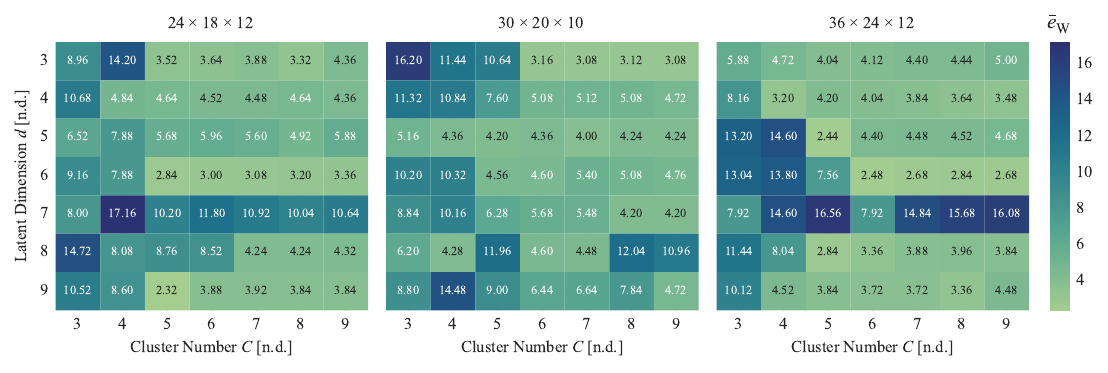}
         \caption{Weighted Misassignment Rate.}
     \end{subfigure}
     \hfill
     \begin{subfigure}[b]{\textwidth}
         \centering
         \includegraphics[width=\textwidth]{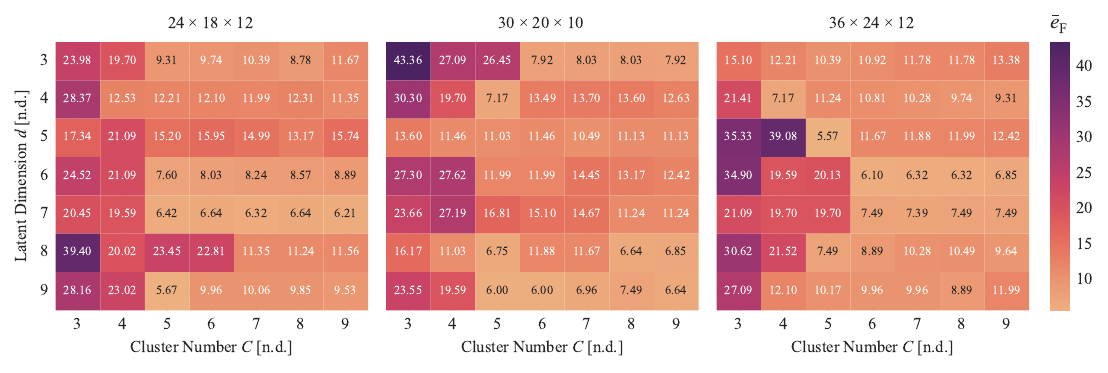}
         \caption{Failure-Only Weighted Misassignment Rate.}
     \end{subfigure}
     \caption{Misassignment rates of GMVAE hyperparameter sweep.}
     \label{fig:architecture_sweep}
\end{figure}

The weighted and failure-only weighted misassignment rates are shown for each architecture in \cref{fig:architecture_sweep}. Each GMVAE was evaluated on 2,500 unseen Monte Carlo samples from the same distribution as the training data. There is no clear relation between classification accuracy and architecture size. The best-performing GMVAE in terms of weighted misassignment is the $d=9, C=5$ GMVAE with the $24 \times 18 \times 12$ architecture, with a weighted misassignment rate of only 2.32\%, and the best-performing GMVAE in terms of failure-only weighted misassignments is the $d=5, C=5$ GMVAE with the $36 \times 24 \times 12$ architecture, with a failure-only weighted missassignment rate of 5.57\%. 

%These results suggest that the selected GMVAE models are sufficiently accurate to serve as probabilistic indicator functions within the aerocapture guidance algorithm. 

\begin{figure}[htbp]
    \centering
    \begin{subfigure}[b]{0.49\linewidth}
        \centering
        \includegraphics[width=\textwidth]{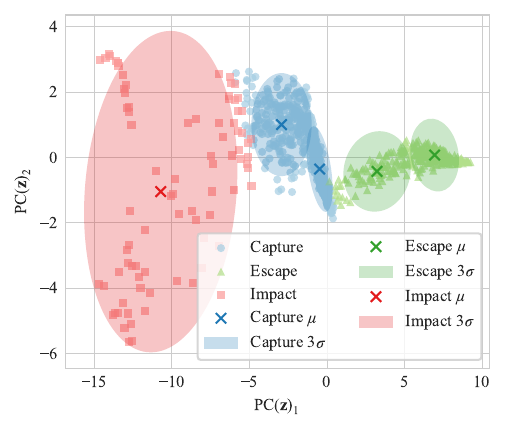}
        \caption{$d=9, C=5$ GMVAE with the $24 \times 18 \times 12$ architecture.}
        \label{fig:pca_weighted}
    \end{subfigure}
    \hfill
    \begin{subfigure}[b]{0.49\linewidth}
        \centering
        \includegraphics[width=\textwidth]{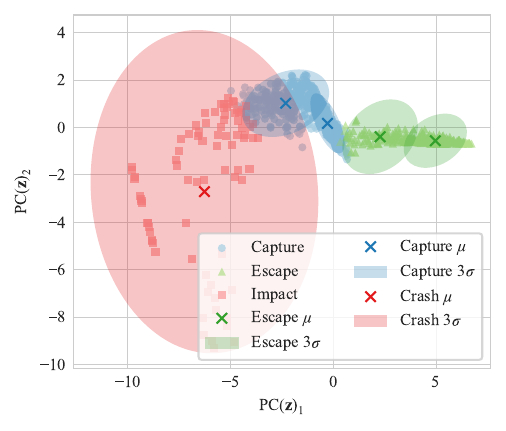}
        \caption{$d=5, C=5$ GMVAE with the $36 \times 24 \times 12$ architecture.}
        \label{fig:pca_55}
    \end{subfigure}
    \caption{Principal component analysis of the GMVAE latent space for well-performing models.}
    \label{fig:latent_samples}
\end{figure}

Principal component analysis (PCA) of first two modes of the latent space shows that the GMVAE clusters agree with the truth mappings, as seen in \cref{fig:latent_samples}. The data points are colored by their true cluster membership. While only the first two modes of the five- or nine-dimensional latent spaces are presented here, the absent modes also show good agreement between the data points and cluster means and variances. The latent spaces are clearly non-Gaussian, and training GMVAE with more than the prescribed three clusters yields a more flexible model that better describes the data. These mappings also reveal the cause of the misclassifications seen above. For example, some capture datapoints fall within the impact cluster $3\sigma$ for the $d=5, C=5$ GMVAE. While further hyperparameter tuning could improve the misclassification rates, the current GMVAE models are sufficient for the probabilistic indicator function, as guidance parameters can be adjusted to account for the misclassifications. 

\section{$\pi$PAG: Probabilistic Indicator Function Predictor-Corrector Aerocapture Guidance Algorithm}

While \eFNPAG is a robust guidance algorithm, all guidance decisions are made based on the onboard mean state. Discrepancies between the onboard guidance and truth models can lead to suboptimal guidance decisions, especially when the vehicle is near the boundary between successful capture and failure (i.e. escape or impact). The use of a probabilistic indicator allows the guidance algorithm to account for the differences in these models and execute guidance decisions to increase the likelihood of successful capture. The GMVAE is particularly suitable for this task because it can efficiently cluster trajectory data and provide accurate modal probabilities.

\subsection{Probabilistic Indicator Function Evaluation}

The GMVAE training algorithm employed in this paper uses an EM step during training to estimate cluster probabilities, and the expectation portion of that algorithm is leveraged within the guidance algorithm to determine the cluster membership probabilities for a given trajectory~\cite{fan_physically_2025}. To find the cluster membership probabilities for sample $\mathbf{x}$, posterior probability $p(c|\mathbf{z})$ is evaluated via Bayes rule:
\begin{equation}
    p(c|\mathbf{z}) = \frac{p(\mathbf{z}|c)p(c)}{p(\mathbf{z})},
\end{equation}
where $\mathbf{z}$ is the latent encoding of $\mathbf{x}$, and the prior probability of cluster $c$ is $p(c) = \pi_c$. The likelihood of the latent sample $\mathbf{z}$ under cluster $c$ is given by $p(\mathbf{z}|c) = \mathcal{N}(\mathbf{z}|\boldsymbol{\mu}_c, \boldsymbol{\sigma}^2_c)$. The marginal probability of $\mathbf{z}$ is $p(\mathbf{z}) = \sum_k \pi_k \mathcal{N}(\mathbf{z}|\boldsymbol{\mu}_k, \boldsymbol{\sigma}^2_k)$.

The membership probabilities for a sample $\mathbf{x}$ are computed in the expectation step as follows:
\begin{enumerate}
    \item {Encode} the sample $\mathbf{x}$ to get the mean $\boldsymbol{\mu}$ and log variance $\log \boldsymbol{\sigma}^2$ of the latent distribution $q(\mathbf{z}|\mathbf{x})$, the approximate posterior over $\mathbf{z}$.
    \item {Compute the Log Probabilities of Each Cluster}: 
    \begin{equation} \label{eqn:log_pz_c}
        \log p(\mathbf{z},c) = \log \pi_c + \log \mathcal{N}(\mathbf{z}|\boldsymbol{\mu}_c, \boldsymbol{\sigma}^2_c).
    \end{equation}
    \item {Apply Bayes rule to obtain the posterior $p(c|\mathbf{z})$}:
    \begin{equation} \label{eqn:gamma_c}
        \gamma_c = p(c|\mathbf{z}) = \frac{p(\mathbf{z},c)}{\sum_k p(\mathbf{z},k)}.
    \end{equation}
\end{enumerate}

As the nominal trajectory is already integrated forward in time during the prediction step of the \eFNPAG algorithm, the only additional operations to use the GMVAE indicator function are (1) encoding the predicted trajectory through the neural network and (2) evaluating the cluster membership probabilities for the predicted trajectory with the above steps. These operations are computationally efficient and can be done in real-time. The probabilistic indicator function is incorporated into guidance by evaluating Algorithm \ref{alg:gmvae_guidance} after the NPC and lateral logic on line \ref{alg:gmvae_insert} in Algorithm \ref{alg:fnpag}. When Algorithm \ref{alg:fnpag} and \ref{alg:gmvae_guidance} are used in conjunction, it is referred to as Probabilistic Indicator Predictor-Corrector Aerocapture Guidance ($\pi$PAG). 

\begin{algorithm}[htb]
	\caption{Probabilistic Indicator Function Evaluation in \piPAG}
	\label{alg:gmvae_guidance}
	\KwIn{Current time $t_k$, predicted trajectory $\mathbf{x}_{t_{k+1}:t_f}$, trajectory history $\mathbf{x}_{t_0:t_k}$, GMVAE encoder \texttt{encoder}, learned cluster probabilities $\pi_c$, learned cluster means $\boldsymbol{\mu}_c$, learned cluster variances $\boldsymbol{\sigma}_c$, capture and failure probability thresholds $\epsilon_C$ and $\epsilon_F$, computed guidance command $\sigma^*$, guidance phase \texttt{phase}, bank correction magnitude $\sigma'$, persistence time $\tau$, last bank correction time $t_{\text{corr}}$}
	\KwOut{Corrected guidance command $\hat{\sigma}^{*}$}
	\Comment{Combined as-flown and predicted trajectory to construct encoder input}
	Combine trajectory history and prediction to get $\mathbf{x}_{t_0:t_f}$, downsample to 36 predetermined non-uniform-in-time indices, compute trajectory energy at these points, and normalize to obtain GMVAE input $\mathbf{x}$\;
	\Comment{Evaluated cluster membership probabilities}
	Encode $\mathbf{x}$ to get $\mathbf{z}$ using \texttt{encoder}\;
	Compute $\log p(\mathbf{z},c)$ for each cluster $c$ using $\pi_c$, $\boldsymbol{\mu}_c$, and $\boldsymbol{\sigma}_c$ with \cref{eqn:log_pz_c}, then compute $\gamma_c = p(c|\mathbf{z})$ for each cluster $c$ with \cref{eqn:gamma_c}\;
	\Comment{Apply corrective guidance command}
    \uIf{$\mathbb{P}(\text{capture}) \leq \epsilon_C \quad \vee \quad \mathbb{P}(\text{failure}) \geq \epsilon_F \quad \vee \quad t_k < t_{\text{corr}} + \tau$}{
        \uIf{\texttt{phase} = 1}{
            Force switch to phase 2\;
        }
        \Else{
        \Comment{Correct bank angle towards lift-down if likely to escape.}
        	\uIf{$\mathbb{P}(\text{escape}) > \mathbb{P}(\text{impact})$}{
        		$|\hat{\sigma}^{*}| \gets |\sigma^{*}| + \sigma'$\;
        		\uIf{$|\hat{\sigma}^{*}| > 180^\circ$}{
        			$|\hat{\sigma}^{*}| \gets 180^\circ$\;}
        		$\hat{\sigma}^{*} \gets \text{sign}(\sigma^{*}) \times |\hat{\sigma}^{*}|$\;
        	}
        \Comment{Correct bank angle towards lift-up if likely to impact.}
        	\Else{
        		$|\hat{\sigma}^{*}| \gets |\sigma^{*}| - \sigma'$\;
        		\uIf{$|\hat{\sigma}^{*}| > |\sigma^{*}|$}{
        			$|\hat{\sigma}^{*}| \gets 0^\circ$\;}
        		$\hat{\sigma}^{*} \gets \text{sign}(\sigma^{*}) \times |\hat{\sigma}^{*}|$\;
        	}
	        \uIf{$\mathbb{P}(\text{capture}) \leq \epsilon_C \quad \vee \quad \mathbb{P}(\text{failure}) \geq \epsilon_F$}{
	            $t_{\text{corr}} \gets t_k$\;
	        }
        }   
    }
\end{algorithm}

After encoding the as-flown and predicted trajectory with the GMVAE, the cluster membership probabilities $\gamma_c$ are computed using steps 1) through 3) above. If the capture probability is less than or equal to $\epsilon_C$, or the escape or impact probability is greater than or equal to $\epsilon_F$, a corrective guidance action is applied. If guidance is in phase 1, the algorithm will force a switch to phase 2 to solve for the optimal bank angle. This indicates that the vehicle should already be in the lift-down orientation for near-escape scenarios, or in the lift-up orientation for near-impact scenarios. If the guidance is in phase 2, the bank angle command is corrected by correction factor $\sigma'$ in the direction of the predicted bank angle command $\sigma^*$. If the vehicle has a high escape probability, the bank angle is corrected to be more lift-down, and if the vehicle has a high impact probability, the bank angle is corrected to be more lift-up. When the bank angle is corrected to be more lift-down, the commanded bank angle is clipped at $180^\circ$, and when the bank angle is corrected to be more lift-up, it is clipped at $0^\circ$. 

This algorithm also uses a persistence time $\tau$ to prevent an oscillatory structure in the bank angle command. If it is within $\tau$ seconds of the last bank angle correction at $t_{\text{corr}}$, the bank angle command is still corrected by $\sigma'$ in the direction of the predicted bank angle command $\sigma^*$. This prevents the algorithm from rapidly fluctuating between corrected and uncorrected bank angles, as the guidance correction alters cluster membership probability. In addition, as the GMVAE is probabilistic, the persistence time allows the algorithm to account for the uncertainty in cluster membership probabilities and ensure more trajectories are saved.  

To illustrate the importance of $\tau$, suppose a bank angle correction occurs at $t_1$. At subsequent time $t_2 = t_1 + \delta t$, capture probability is above $\epsilon_C$, and no corrective action is applied. However, since a single time step is insufficient to fully correct the trajectory, the capture probability falls below $\epsilon_C$ at $t_3 = t_2 + \delta t$, necessitating a further corrective adjustment. By enforcing the persistence interval $\tau$, the bank angle remains partially corrected even when the thresholds are temporarily satisfied, preventing rapid on-off corrections.

While using a fixed bank correction magnitude and persistence time is an effective way to implement the probabilistic indicator function, this approach is not necessarily the only or optimal method for aerocapture guidance. Other schemes\textemdash such as a variable bank angle correction magnitude or persistence time based on the cluster membership probabilities\textemdash could be implemented to improve performance. 

\subsection{\piPAG Parameter Tuning}
\label{sec:pipag_tuning}

This implementation of a probabilistic indicator function has three tuning parameters: the failure probability threshold, $\epsilon_F$, the bank correction magnitude, $\sigma'$, and the persistence time, $\tau$. The capture probability threshold, $\epsilon_C$, is computed as $1 - \epsilon_F$. Smaller bank correction magnitudes and shorter persistence times lead to more conservative corrections. If these parameters are too small, \piPAG cannot recover all failed cases. If these parameters are too large, \piPAG can overcorrect, causing a previously escaped case to impact or a previously impacted case to escape. To determine the optimal parameters for \piPAG for this problem, all combinations of $\epsilon_F$ values of $\left[ 2.5 \times 10^{-2}, 2.0 \times 10^{-2}, 1.5 \times 10^{-2}, 1.0 \times 10^{-2}, 1.0 \times 10^{-3}, 1.0 \times 10^{-4}, 1.0 \times 10^{-5}, 1.0 \times 10^{-6} \right]$, $\sigma'$ values of $\left[ 10,  20, 30, 40, 50, 60,70, 80\right]^\circ$, and $\tau$ values of $\left[ 10,  20, 30, 40, 50, 60,70, 80\right]$ s were evaluated for the recoverable failure cases. To determine a suitable combination for this study, 2,500 samples from the GMVAE training distribution were evaluated with \eFNPAG, and each parameter combination was tested in \piPAG for the 24 recoverable cases (12 escapes and 12 impacts). 

\begin{figure}[htbp]
    \centering
    \begin{subfigure}[b]{0.49\linewidth}
        \centering
        \includegraphics[width=\textwidth]{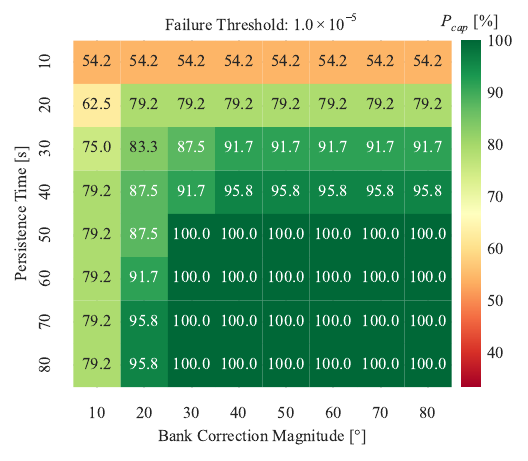}
        \caption{Heatmap for $\epsilon_F=1.0e-5$.}
        \label{fig:sweep_heatmap}
    \end{subfigure}
    \hfill
    \begin{subfigure}[b]{0.49\linewidth}
        \centering
        \includegraphics[width=\textwidth]{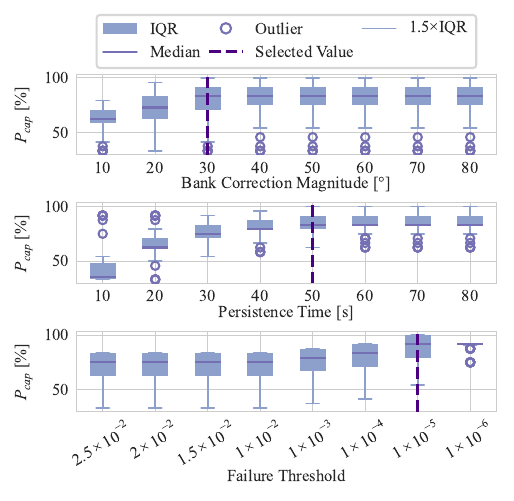}
        \caption{Marginalized parameter sweep results.}
        \label{fig:sweep_marginals}
    \end{subfigure}
    \caption{Results of \piPAG tuning parameter sweep.}
    \label{fig:sweep}
\end{figure}

\cref{fig:sweep} shows the results of the parameter sweep. \cref{fig:sweep_marginals} depicts a boxplot of the capture percentage, $\mathbb{P}_C$, for marginalized distributions of each tuning parameter. For example, the data plotted for a $\sigma'=10^\circ$ includes all possible combinations of $\epsilon_F$ and $\tau$ for that bank correction magnitude. The selected values for each parameter are shown in dashed lines\textemdash beyond these values, there is no improvement in capture rate. This can also be seen in the heatmap plot for $\epsilon_F = 1.0\times10^{-5}$ in \cref{fig:sweep_heatmap}. Beyond bank correction magnitudes of 30$^\circ$ and persistence times of 50 s, all cases are recovered. For the failure threshold, there is a decrease in performance past $1.0\times10^{-5}$. This occurs because the GMVAE assigns a minimum probability of $1.0\times10^{-6}$ to each cluster during encoder evaluation to avoid zero cluster probabilities (see \cref{sec:gmvae_training_nn}). If the failure threshold is set at $1.0\times10^{-6}$, corrective guidance commands are taken at every timestep as failure probability always exceeds the failure threshold by definition. This leads to some previously-escaped trajectories impacting and some previously-impacted trajectories escaping. Thus, the parameters used for this implementation of \piPAG are a failure threshold of $\epsilon_F=1.0\times10^{-5}$, a bank correction magnitude of $\sigma'=30^\circ$, and a persistence time of $\tau=50$ seconds.

\subsection{Representative Example of \piPAG Guidance Behavior}

To demonstrate how the \piPAG algorithm can save otherwise failed \eFNPAG trajectories, the performance of \eFNPAG and \piPAG was compared for a single near-escape scenario. The initial condition and vehicle parameters for this specific scenario are shown in \cref{tab:example_conditions}, and the truth atmosphere model was a GRAM profile with a density perturbation scale of \texttt{dp} = 1.5. When using \eFNPAG, these initial conditions resulted in an escape trajectory; however, \piPAG was able to correct the trajectory and achieve a successful capture. The \piPAG parameters used for this scenario were $\epsilon_F = 1.0 \times 10^{-5}$, $\sigma' = 30^\circ$, and $\tau = 50$ s (see \cref{sec:pipag_tuning} for details on selection). 

\begin{table}[h]
    \centering 
    \caption{Initial conditions and parameter dispersions for off-nominal scenario.}
    \label{tab:example_conditions}
    \begin{tabular}{cccccccc}
\hline
$h_0$ [km] & $V_{0,I}$ [km/s] & $\gamma_{0,I}$ [$^\circ$] & $\psi_{0,I}$ [$^\circ$] & $\theta_0$ [$^\circ$] & $\phi_0$ [$^\circ$] & m [kg] & $C_L$ \\
\hline
1083.00 & 26.33 & -9.87 & 45.04 & 189.88 & -9.62 & 2444.0 & 0.28 \\
\hline
\end{tabular}
\end{table}

\begin{figure}[hbt]
	\centering
	\begin{subfigure}[b]{0.49\textwidth}
         \centering
         \includegraphics[width=\textwidth]{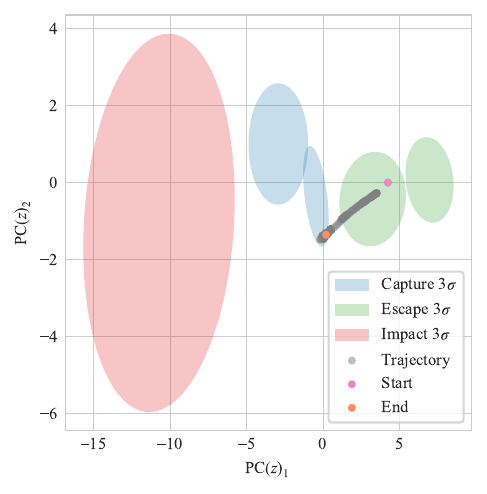}
         \caption{PCA of trajectory in latent space.}
         \label{fig:pca_single_example}
     \end{subfigure}
     \hfill
     \begin{subfigure}[b]{0.49\textwidth}
         \centering
         \includegraphics[width=\textwidth]{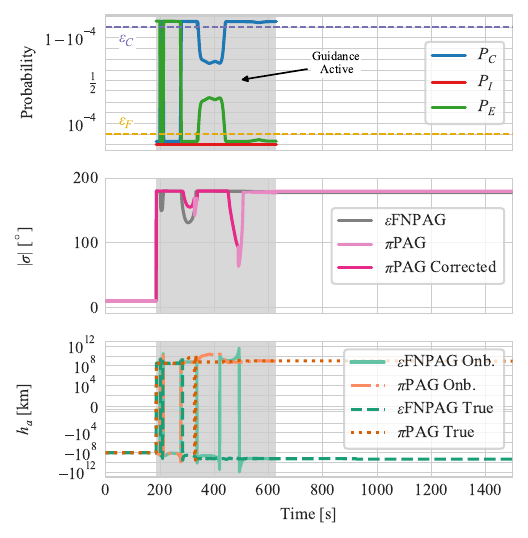}
         \caption{Trajectory performance over time.}
         \label{fig:single_example_perf}
     \end{subfigure}
     \caption{\piPAG and \eFNPAG latent space trajectory, cluster membership probability, bank angle command, and predicted apoapsis altitude for illustrative example.}
     \label{fig:latent_space_escape}
\end{figure}

\cref{fig:latent_space_escape} shows the trajectory evolution in the latent space, cluster membership probabilities, bank angle magnitudes, and onboard and truth predicted apoapsis for both \eFNPAG and $\pi$PAG. The \piPAG bank angle commands are colored based on whether a correction is applied: dark pink indicates corrected commands---corresponding to high failure probability---while light pink indicates uncorrected commands---corresponding to high capture probability. Initially, the trajectory has a high escape probability and is near the escape clusters in latent space. As the guidance applies corrective bank angles (shown in dark pink), the escape probability decreases while the capture probability increases. Compared to $\varepsilon$FNPAG, \piPAG commands a bank angle magnitude closer to 180$^\circ$ early in the trajectory, flying more lift-down and resulting in a successful capture. This outcome is reflected in the terminal position in latent space, which lies in one of the capture clusters.

Because the indicator function predicted that this scenario was likely to escape, the \piPAG algorithm was able to correct the trajectory and achieve a successful capture, even as the onboard guidance model did not account for true environment dispersions. This can be seen in the apoapsis altitude plot in \cref{fig:single_example_perf}: midway through the guided portion of the trajectory (approximately 300-350 seconds), there is disagreement between the truth and onboard predicted apoapsis for $\varepsilon$FNPAG. The onboard model (light green line) predicts a positive apoapsis radius, while the truth model (dark green dashed line) shows a negative apoapsis radius, indicating escape. During this period, \eFNPAG did not command full lift-down, and even this small region of partial lift-up resulted in an escaped trajectory. During the same time period, the truth and onboard models for \piPAG agreed that the trajectory had successfully captured. 

 The apoapsis error and required propellant for the \piPAG trajectory were approximately 416,145 km and 301.83 m/s, respectively. While the apoapsis error was large, the required $\Delta v$ was within the expected range for proposed ice giant aerocapture missions (and still significantly lower than that required for a propulsive orbit insertion maneuver) \cite{deshmukh_performance_2024}.

This case highlights several noteworthy features of the \piPAG implementation. First, it demonstrates the value of incorporating a persistence time into the algorithm. The persistence time allows the algorithm to smooth the bank angle command and prevent rapid oscillations between corrected and uncorrected bank angles. Second, it demonstrates the flexibility provided through tuning $\epsilon_F$ and $\epsilon_C$. If these thresholds were set less conservatively, corrective guidance would not be taken during periods where the capture probability is moderate (e.g., approximately 350-450 seconds, with capture probability near 60\% and escape probability near 40\%). Third, the probabilistic nature of the GMVAE indicator function enables accurate classification of trajectory modes using both as-flown and predicted trajectory data. Even when the onboard model differs from the truth model, the GMVAE provides reliable cluster membership probabilities, demonstrating the algorithm’s robustness under model mismatch.

\section{Performance Evaluation Against State-of-the-Art Aerocapture Guidance}
\label{sec:results}

The proposed \piPAG guidance algorithm is evaluated against the $\varepsilon$FNPAG algorithm across a range of representative aerocapture scenarios. Four applications of \piPAG are presented. First, the performance of \piPAG is compared to \eFNPAG for samples within the GMVAE training distribution and atmosphere model. Next, \piPAG is shown to be generalizable to initial condition distributions and atmosphere models outside the GMVAE training data. Then, the use of a probabilistic indicator function is compared with the use of a first-order fading memory filter for density estimation in aerocapture guidance. Finally, the performance of different GMVAEs as probabilistic indicator functions is evaluated by comparing the best GMVAE in terms of weighted misassignment rate to the best GMVAE in terms of failure-only weighted misassignment rate. 

The following results are generated using high-fidelity sampling and evaluated using nonparametric statistical tests to assess significance. Latin Hypercube Sampling (LHS) was employed to ensure the samples are a good representation of the real variability in the distribution~\cite{iman_small_1980}. A Mann-Whitney U test was used to determine if the change the apoapsis radius distribution or $\Delta v$ distribution is significant for the following results. This statistical test determines if the underlying distribution between sets of independent samples are the same, without assuming the form of the underlying distributions~\cite{macfarland_mannwhitney_2016}. The result of the Mann-Whitney U test is a $p$ value. A $p$ value less than 0.05 indicates that there is less than a 5\% probability of observing the measured difference between distributions if they were in fact the same, and we therefore conclude the distributions are significantly different.

\subsection{\piPAG Performance Within Training Conditions}
\label{sec:samedist_samedp}

A 2,500-sample Monte Carlo simulation was conducted using the same samples for both $\varepsilon$FNPAG and \piPAG from an initial GU Mixture of initial conditions and vehicle parameters. The center point and $3\sigma$ values for each variable are the same as used for GMVAE training and are shown in \cref{tab:gmave_training_mean}. The truth atmosphere model is GRAM with a \texttt{dp} = 1.5. 

\piPAG increases capture rate and decreases failure rate for samples within the GMVAE training conditions, as shown in \cref{tab:samedist_samedp_outcome}. \piPAG saves all recoverable escape and impact cases. 

\begin{table}[htb]
    \centering 
    \caption{\piPAG and \eFNPAG outcome comparison for captured cases for conditions within GMVAE training conditions.}
    \label{tab:samedist_samedp_outcome}
    \begin{tabular}{lccc>{\centering\arraybackslash}p{2cm}>{\centering\arraybackslash}p{2cm}}
	\hline
	 & Capture \% & Impact \% & Escape \% & Recoverable Escape \% & Recoverable Impact \% \\
	\hline
	$\varepsilon$FNPAG & 62.32 & 7.64 & 30.04 & 0.00 & 0.00 \\
	$\pi$PAG & 63.28 & 7.16 & 29.56 & 100.00 & 100.00 \\
	\hline
	\end{tabular}	
\end{table}

\begin{figure}[hbt]
	\centering
	\begin{subfigure}[b]{0.49\textwidth}
         \centering
         \includegraphics[width=\textwidth]{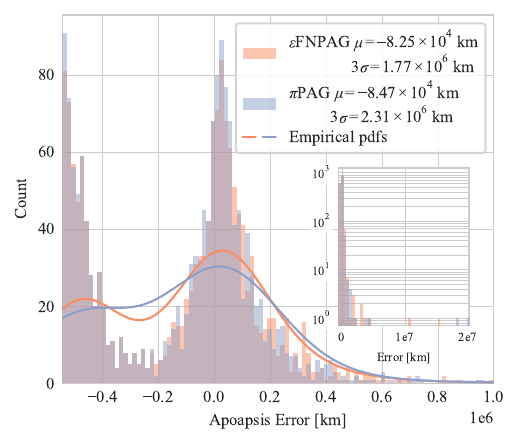}
         \caption{Apoapsis Error. Inset plot shows outliers.}
         \label{fig:samedist_samedp_apo}
     \end{subfigure}
     \hfill
     \begin{subfigure}[b]{0.49\textwidth}
         \centering
         \includegraphics[width=\textwidth]{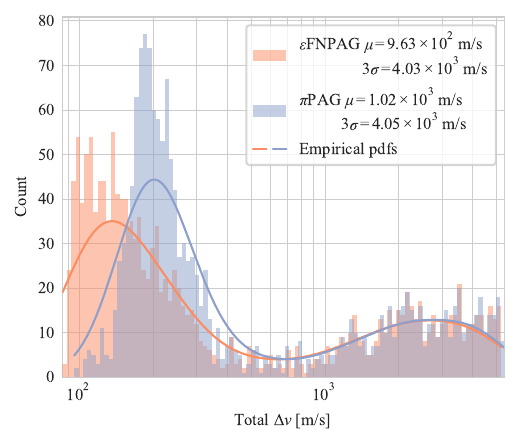}
         \caption{$\Delta v$.}
         \label{fig:samedist_samedp_dv}
     \end{subfigure}
     \caption{\piPAG and \eFNPAG performance comparison for GU Mixture Initial Conditions and GRAM truth atmosphere model with \texttt{dp} = 1.5.}
     \label{fig:samedist_samedp_perf}
\end{figure}

For the captured cases, \piPAG has little change in apoapsis error, but does significantly increase the $\Delta v$ required, as shown in \cref{fig:samedist_samedp_perf}. The empirical pdfs on \cref{fig:samedist_samedp_perf} are generated using kernel density estimation with the Python \texttt{seaborn} package~\cite{Waskom2021}. Although \piPAG has slightly higher apoapsis error mean and standard deviation for the captured cases, the difference in apoapsis error is not statistically significant with a Mann-Whitney U test $p$ value of 0.414. The increase in apoapsis standard deviation is due to the recovery of highly elliptical cases, which result in outlier apoapsis radii shown in the inset in \cref{fig:samedist_samedp_apo}. The increase in required $\Delta v$, however, is statistically significant between the two models at a $p$ value of zero. 

\subsection{\piPAG Generalization Beyond Training Conditions}
\label{sec:diff_conds}

A common concern with data-driven deep learning methods is their limited generalizability to data outside the training set. To determine if the GMVAE is generalizable to data outside its training dispersions, Monte Carlo simulations were conducted for both distributions and atmosphere models outside the training data.

\subsubsection{\piPAG Robustness to Differing Initial Condition Distributions}
\label{sec:diffdist_samedp}

\eFNPAG and \piPAG Monte Carlo simulations were evaluated for initial conditions sampled from a Student's $t$ distribution with $\nu=3$, a near-escape Gaussian distribution, and a near-impact Gaussian distribution. All samples were drawn using LHS. The mean and variance parameters of these distributions are shown in \cref{tab:diffdist_samedp_dispersions}. The truth atmosphere model for these simulations is a GRAM atmosphere with a \texttt{dp} = 1.5. 

Outer planet aerocapture missions generally target high apoapsis radius orbits to reduce the risk of impact~\cite{deshmukh_performance_2024}, and large dispersions in initial state and vehicle parameters will lead to some escaped trajectories when targeting the middle of the flight path angle corridor. The Student's $t$ and near-escape Gaussian scenarios used an apoapsis target altitude of 550,000 km and a periapsis target of 4,000 km, while the near-impact Gaussian scenario used an apoapsis target altitude of 100,000 km and a periapsis target altitude of 3,000 km. The lower apoapsis and periapsis targets encourage more impact trajectories. Although targeting a low apoapsis orbit is generally avoided in mission design to reduce the risk of impact, a scenario with a steep entry flight path angle and low apoapsis radius was intentionally chosen to demonstrate $\pi$PAG's ability to recover from potential impact trajectories. These test cases evaluated $\pi$PAG's ability to recover from potential impacts, a capability that could be critical in environments with lower-density atmospheres.

\begin{table}[htb]
    \centering 
    \caption{Monte Carlo Initial Conditions for Differing Distributions.}
    \label{tab:diffdist_samedp_dispersions}
        \begin{tabular}{@{}p{1.8cm}lcccccccc@{}}
    \hline
       Distribution & Parameter & $h_0$ [km] & $\theta_0$ [$\deg$] & $\phi_0$ [$\deg$] & $V_{0,I}$ [km/s] & $\gamma_{0,I}$ [$\deg$] & $\psi_0$ [$\deg$] &  $\frac{L}{D}$ [n.d.] & $m$ [kg] \\ \hline
      {Student's $t$,} & $\mu$ & 1000 & 190.045 & -9.764 & 24.936 & -10.572 & 45.00 & 0.25 & 2847.068 \\ 
      $\nu=3$ & $3\sigma$ & 100 & 0.227 & 0.116 & 0.750 & 0.50 & 0.063 & 0.075 & 854.120 \\
      \hline
      {Gaussian} & $\mu$ & 1000 & 190.045 & -9.764 & 24.936 & -10.572 & 45.00 & 0.25 & 2847.068 \\ 
      Near-Escape & $3\sigma$ & 100 & 0.227 & 0.116 & 0.750 & 0.50 & 0.063 & 0.075 & 854.120 \\
      \hline
      {Gaussian} & $\mu$ & 1000 & 190.045 & -9.764 & 24.936 & -11.278 & 45.00 & 0.25 & 2847.068 \\ 
      Near-Impact & $3\sigma$ & 100 & 0.227 & 0.116 & 0.750 & 0.50 & 0.063 & 0.075 & 854.120 \\
      \hline
    \end{tabular}
\end{table}

To draw a scalar sample $x_i$ from an independent Student's $t$ distribution with $\nu$ degrees of freedom, we compute $x_i = \mu_i + \sigma_i \cdot t^{-1}(U_i,\, \nu)$, where $\mu_i$ and $\sigma_i$ are the desired mean and standard deviation of the $i$-th parameter, $U_i \sim \mathcal{U}(0,1)$ is a uniform sample drawn via LHS, and $t^{-1}(\cdot,\, \nu)$ denotes the inverse CDF of the Student's $t$ distribution with $\nu$ degrees of freedom. Compared to a Gaussian distribution, the Student's $t$ distribution has heavier tails. As $\nu \rightarrow \infty$, the Student's $t$ distribution converges to a standard normal distribution, and so the choice of $\nu$ controls the degree to which outliers are represented in the sample set.

\begin{figure}[hbt]
	\centering
	\begin{subfigure}[b]{0.49\textwidth}
         \centering
         \includegraphics[width=\textwidth]{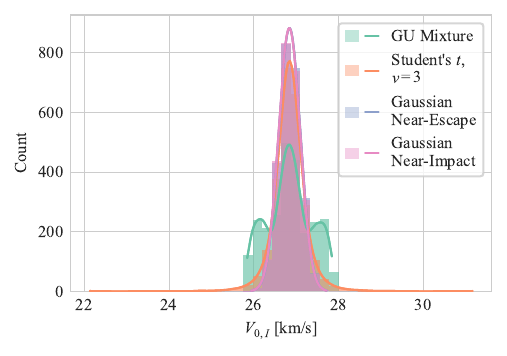}
         \caption{Velocity.}
     \end{subfigure}
     \hfill
     \begin{subfigure}[b]{0.49\textwidth}
         \centering
         \includegraphics[width=\textwidth]{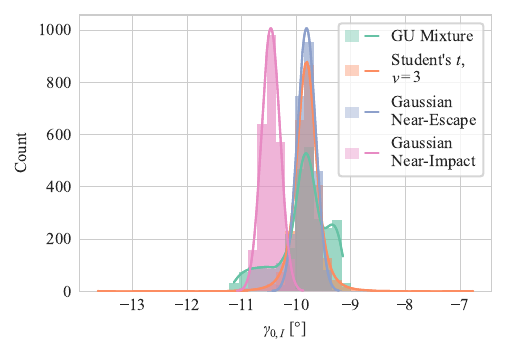}
         \caption{Flight Path Angle.}
     \end{subfigure}
     \caption{Initial condition samples for differing input distributions.}
     \label{fig:diffdist}
\end{figure}

The samples from these distributions in velocity and flight path angle are compared with the GMVAE training data distribution in \cref{fig:diffdist}. The negative bias in the lower tail of the GU mixture for flight path angle encompasses the near-impact Gaussian flight path angle samples, while the outliers from the Student's $t$ distribution are not within the training data. 

The outcomes of these trials are shown in \cref{tab:diffdist_samedp_outcome}. \piPAG increases capture rate over $\varepsilon$FNPAG, saving 100\% of recoverable impact cases for both the Student's $t$ and Gaussian near-impact scenarios. Between 71.43 and 79.31\% of recoverable escaped trajectories are saved for the Student's $t$ and Gaussian near-escape cases: while corrective guidance was taken for the recoverable trajectories which \piPAG failed to save, it was not sufficient to change the outcome. 

\begin{table}[htb]
    \centering 
    \caption{\piPAG and \eFNPAG outcome comparison for distributions outside GMVAE training conditions.}
    \label{tab:diffdist_samedp_outcome}
	\begin{tabular}{llccc>{\centering\arraybackslash}p{2cm}>{\centering\arraybackslash}p{2cm}}
	\hline
	Distribution & Guidance & Capture \% & Impact \% & Escape \% & Recoverable Escape \% & Recoverable Impact \% \\
	\hline
	\multirow{2}{*}{Student's $t$, $\nu=3$} & $\varepsilon$FNPAG & 77.64 & 1.76 & 20.60 & 0.00 & 0.00 \\
	 & $\pi$PAG & 78.40 & 1.60 & 20.00 & 71.43 & 100.00 \\
	\hline
	\multirow{2}{*}{Gaussian Near-Escape} & $\varepsilon$FNPAG & 86.04 & 0.00 & 13.96 & 0.00 & N/A \\
	 & $\pi$PAG & 86.96 & 0.00 & 13.04 & 79.31 & N/A \\
	\hline
	\multirow{2}{*}{Gaussian Near-Impact} & $\varepsilon$FNPAG & 94.84 & 5.12 & 0.04 & N/A & 0.00 \\
	 & $\pi$PAG & 95.76 & 4.16 & 0.08 & N/A & 100.00 \\
	\hline
	\end{tabular}
\end{table}

Additionally, for the near-impact scenario, \piPAG results in one more escape than $\varepsilon$FNPAG. This case captured when using $\varepsilon$FNPAG, but escaped because of bank corrections applied in $\pi$PAG. In this scenario, the probabilistic indicator initially predicted a high impact probability, and the bank angle was corrected more lift-up. While the GMVAE predicted high escape probability later on, and \piPAG corrects the bank angle lift-down, this trajectory still escapes. This phenomenon is discussed further in \cref{sec:discussion}. 

The apoapsis error and $\Delta v$ required for each distribution and guidance algorithm are shown in \cref{tab:diffdist_samedp_perf}. Again, the difference in apoapsis error is not statistically significant, with a Mann-Whitney U test $p$ values of 0.352 for the Student's $t$, 0.371 for the Gaussian near-escape, and 0.723 for the Gaussian near-impact. Conversely, the increase in required $\Delta v$ is statistically significant with a $p$ value of zero for all distributions.  

\begin{table}[hbt]
\centering
\caption{$\pi$PAG and $\varepsilon$FNPAG performance comparison for captured cases for distributions outside GMVAE training conditions.}
\label{tab:diffdist_samedp_perf}
\begin{tabular}{llcccc}
\hline
Distribution & Guidance & \begin{tabular}{@{}c@{}} $r_a$ Mean \\ $[\times 10^{4}$ km$]$\end{tabular} & \begin{tabular}{@{}c@{}}$r_a$ 3$\sigma$ \\ $[\times 10^{5}$ km$]$ \end{tabular} & \begin{tabular}{@{}c@{}}$\Delta v$ Mean  \\ $[\times 10^{2}$ m/s$]$ \end{tabular} & \begin{tabular}{@{}c@{}}$\Delta v$ 3$\sigma$ \\ $[\times 10^{3}$ m/s$]$ \end{tabular} \\
\hline
\multirow{2}{*}{Student's T, $\nu=3$} & $\varepsilon$FNPAG & 0.98 & 13.57 & 4.56 & 2.55 \\
 & $\pi$PAG & -0.64 & 8.48 & 5.05 & 2.55 \\
\hline
\multirow{2}{*}{Gaussian Near-Escape} & $\varepsilon$FNPAG & 6.00 & 12.14 & 2.71 & 1.40 \\
 & $\pi$PAG & 6.06 & 26.71 & 3.19 & 1.36 \\
\hline
\multirow{2}{*}{Gaussian, Near-Impact} & $\varepsilon$FNPAG & -2.68 & 0.97 & 10.41 & 2.89 \\
 & $\pi$PAG & -2.47 & 1.24 & 11.93 & 2.73 \\
\hline
\end{tabular}
\end{table}

\subsubsection{\piPAG Robustness to Alternative Atmosphere Models}
\label{sec:diffdist_diffdp}

The GMVAE used in \piPAG was trained on \eFNPAG simulations using a GRAM truth atmosphere model with a \texttt{dp} = 1.5 and an initial GU Mixture distribution. In \cref{sec:diffdist_samedp}, we presented \piPAG performance results for distributions outside the GMVAE training data. In this section, we present results for both distributions and truth atmosphere models outside of the GMVAE training data.  

2,500 Monte Carlo samples are drawn from the near-escape and near-impact Gaussian distributions described in \cref{tab:diffdist_samedp_dispersions} using LHS. Truth atmosphere samples are taken from GRAM with a \texttt{dp} = 2. A comparison of the GRAM atmosphere models with varying \texttt{dp} is shown in \cref{fig:atmos}; a \texttt{dp} = 2 introduces significantly more atmospheric density variation than present in the \texttt{dp} = 1.5 profiles used in training data. 

The resultant outcomes are shown in \cref{tab:diffdist_diffdp_outcome}. \piPAG again increases capture rate for both distributions. Similar to the results for \texttt{dp} = 1.5, \piPAG saves all recoverable impact cases in the near-impact Gaussian scenario with increased atmospheric variance, while \piPAG fails to recover 24.59\% of recoverable escape cases in the near-escape Gaussian scenario. We again see that a single case which captured when using \eFNPAG escaped when using $\pi$PAG. This case escapes for the same reasons discussed in \cref{sec:diffdist_samedp}. 

\begin{table}[htb]
    \centering 
    \caption{\piPAG and \eFNPAG outcome comparison for distributions and atmosphere models outside GMVAE training conditions.}
    \label{tab:diffdist_diffdp_outcome}
	\begin{tabular}{p{3.5cm}lccc>{\centering\arraybackslash}p{2cm}>{\centering\arraybackslash}p{2cm}}
	\hline
	Scenario & Guidance & Capture \% & Impact \% & Escape \% & Recoverable Escape \% & Recoverable Impact \% \\
	\hline
	Gaussian Near-Escape & $\varepsilon$FNPAG & 84.84 & 0.00 & 15.16 & 0.00 & N/A \\
	(\texttt{dp} = 2) & $\pi$PAG & 86.64 & 0.00 & 13.36 & 75.41 & N/A \\
	\hline
	Gaussian Near-Impact & $\varepsilon$FNPAG & 95.04 & 4.92 & 0.04 & N/A & 0.00 \\
	(\texttt{dp} = 2) & $\pi$PAG & 95.76 & 4.16 & 0.08 & N/A & 100.00 \\
	\hline
	\end{tabular}
\end{table}

\piPAG results in an increase in apoapsis error mean and standard deviation as seen in \cref	{tab:diffdist_diffdp_perf}, which is not significant with a Mann-Whitney U test $p$ value of 0.096 for the near-escape results and 0.661 for the near-impact results. There is again an increase in mean $\Delta v$ required, which is statistically significant with a $p$ value of zero for both distributions.

\begin{table}[hbt]
	\centering
    \caption{\piPAG and \eFNPAG performance comparison for captured cases for distributions and atmosphere models outside GMVAE training conditions.}
    \label{tab:diffdist_diffdp_perf}
\begin{tabular}{llcccc}
\hline
Scenario & Guidance & \begin{tabular}{@{}c@{}} $r_a$ Mean \\ $[\times 10^{4}$ km$]$\end{tabular} & \begin{tabular}{@{}c@{}}$r_a$ 3$\sigma$ \\ $[\times 10^{5}$ km$]$ \end{tabular} & \begin{tabular}{@{}c@{}}$\Delta v$ Mean  \\ $[\times 10^{2}$ m/s$]$ \end{tabular} & \begin{tabular}{@{}c@{}}$\Delta v$ 3$\sigma$ \\ $[\times 10^{3}$ m/s$]$ \end{tabular} \\
\hline
Gaussian Near-Escape & $\varepsilon$FNPAG & 15.73 & 41.15 & 2.97 & 1.41 \\
(\texttt{dp} = 2) & $\pi$PAG & 18.78 & 82.12 & 3.42 & 1.36 \\
\hline
Gaussian Near-Impact & $\varepsilon$FNPAG & -2.71 & 0.99 & 10.68 & 2.89 \\
(\texttt{dp} = 2) & $\pi$PAG & -2.39 & 1.49 & 12.13 & 2.71 \\
\hline
\end{tabular}
\end{table}

\subsection{Comparison with Fading-Memory Filter Density Estimation}
\label{sec:ff}

In the previous sections, \piPAG was shown to improve capture rate over $\varepsilon$FNPAG. Apoapsis mean error and standard deviation generally increased, but this change was not statistically significant. These prior analyses were performed using first-order fading memory filters for density estimation (\cref{eqn:drag_ff,eqn:lift_ff}), which have been shown to improve aerocapture guidance performance in the presence of uncertainty \cite{lu_optimal_2015, brunner_skip_2008}. These filters improve performance by accounting for the difference between the onboard atmosphere model (here, the polynomial model) and the true atmosphere (a GRAM sample). Because density is a major source of uncertainty in aerocapture, improving density knowledge with a fading filter can improve overall aerocapture performance. In contrast, the probabilistic indicator function is not directly designed to improve apoapsis targeting accuracy by improving density estimation. Rather, it is designed to improve the aerocapture success rate by correcting trajectories that would otherwise escape or impact. These two augmentations to NPC aerocapture guidance are compared below.

A 2,500-sample Monte Carlo simulation was run for the near-escape and near-impact scenario using \eFNPAG with and without a fading memory filter and \piPAG with and without a fading memory filter. The samples were drawn from the distributions described in \cref{tab:diffdist_samedp_dispersions} using LHS. These simulations were run with truth GRAM atmosphere profiles with \texttt{dp} = 2 for the near-escape scenario and \texttt{dp} = 1.5 for the near-impact scenario.

Outcome results are shown in \cref{tab:ff_outcomes}. There is slight improvement in capture rate when using a fading filter in both \eFNPAG and \piPAG for the near-escape scenario. However, for the near-impact scenario, there is no improvement in capture rate for \eFNPAG with or without the fading filter. This is likely due to the better match between the polynomial atmosphere model used onboard and the GRAM truth atmosphere model at low altitudes flown through in the near-impact trajectories (see \cref{fig:atmos}). 

\begin{table}[htb]
    \centering
    \caption{\piPAG and \eFNPAG outcome comparison with and without a fading filter for density estimation. }
    \label{tab:ff_outcomes}
	\begin{tabular}{p{2cm}lccc>{\centering\arraybackslash}p{2cm}>{\centering\arraybackslash}p{2cm}}
	\hline
	Scenario & Guidance & Capture \% & Impact \% & Escape \% & Recoverable Escape \% & Recoverable Impact \% \\
	\hline
	\multirow{4}{*}{\shortstack[l]{Gaussian\\Near-Escape\\(\texttt{dp} = 2)}} & $\varepsilon$FNPAG, FF & 84.84 & 0.00 & 15.16 & 0.00 & N/A \\
	 & $\varepsilon$FNPAG, no FF & 84.24 & 0.00 & 15.76 & 0.00 & N/A \\
	 & $\pi$PAG, FF & 86.64 & 0.00 & 13.36 & 75.41 & N/A \\
	 & $\pi$PAG, no FF & 83.96 & 0.00 & 16.04 & 11.54 & N/A \\
	\hline
	\multirow{4}{*}{\shortstack[l]{Gaussian\\Near-Impact\\(\texttt{dp} = 1.5)}} & $\varepsilon$FNPAG, FF & 94.84 & 5.12 & 0.04 & N/A & 0.00 \\
	 & $\varepsilon$FNPAG, no FF & 94.84 & 5.12 & 0.04 & N/A & 0.00 \\
	 & $\pi$PAG, FF & 95.76 & 4.16 & 0.08 & N/A & 100.00 \\
	 & $\pi$PAG, no FF & 95.28 & 4.48 & 0.24 & N/A & 66.67 \\
	\hline
	\end{tabular}
\end{table}

The fading filter also improves recoverable saves when using \piPAG for both scenarios. However, the worst-performing case for the near-escape scenario is \piPAG without a fading filter. We hypothesize that this discrepancy is caused by inaccurate GMVAE classification accuracy for inputs without a fading filter. As the GMVAE is trained on \eFNPAG truth trajectories which employ a fading filter, \piPAG expects its corrective actions to operate in an environment with a fading filter.

The fading filter significantly reduces mean required $\Delta v$ for all scenarios. The pairwise difference in $\Delta v$ between each case in \cref{tab:ff_performance} is statistically significant, with a Mann-Whitney U test $p$ value of less than 0.020 for all comparisons. Thus, the $\Delta v$ savings when using a fading filter (or not using a probabilistic indicator), are significant. 

\begin{table}[h]
    \centering
    \caption{\piPAG and \eFNPAG performance comparison for captured cases with and without a fading filter for density estimation.}
    \label{tab:ff_performance}
\begin{tabular}{p{2cm}lcccc}
\hline
Scenario & Guidance & \begin{tabular}{@{}c@{}} $r_a$ Mean \\ $[\times 10^{4}$ km$]$\end{tabular} & \begin{tabular}{@{}c@{}}$r_a$ 3$\sigma$ \\ $[\times 10^{5}$ km$]$ \end{tabular} & \begin{tabular}{@{}c@{}}$\Delta v$ Mean  \\ $[\times 10^{2}$ m/s$]$ \end{tabular} & \begin{tabular}{@{}c@{}}$\Delta v$ 3$\sigma$ \\ $[\times 10^{3}$ m/s$]$ \end{tabular} \\
\hline
\multirow{4}{*}{\shortstack[l]{Gaussian\\Near-Escape\\(\texttt{dp} = 2)}} & $\varepsilon$FNPAG, FF & 15.73 & 41.15 & 2.97 & 1.41 \\
 & $\varepsilon$FNPAG, no FF & 55.72 & 747.96 & 3.38 & 1.43 \\
 & $\pi$PAG, FF & 18.78 & 82.12 & 3.42 & 1.36 \\
 & $\pi$PAG, no FF & 6.09 & 71.22 & 3.98 & 1.39 \\
\hline
\multirow{4}{*}{\shortstack[l]{Gaussian\\Near-Impact\\(\texttt{dp} = 1.5)}} & $\varepsilon$FNPAG, FF & -2.68 & 0.97 & 10.41 & 2.89 \\
 & $\varepsilon$FNPAG, no FF & -2.75 & 1.00 & 10.64 & 2.85 \\
 & $\pi$PAG, FF & -2.47 & 1.24 & 11.93 & 2.73 \\
 & $\pi$PAG, no FF & -2.54 & 1.52 & 12.05 & 2.68 \\
\hline
\end{tabular}
\end{table} 

\begin{figure}[htbp]
    \centering
    \begin{subfigure}[b]{0.49\linewidth}
        \centering
        \includegraphics[width=\textwidth]{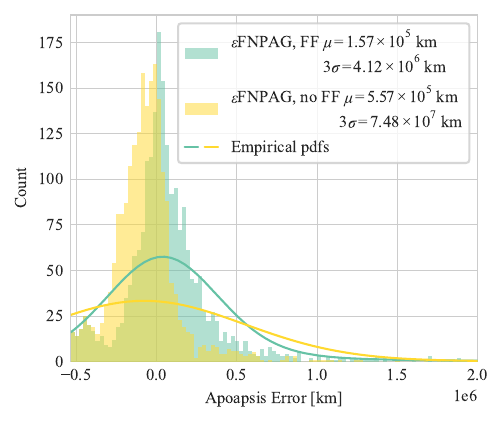}
        \caption{$\varepsilon$FNPAG.}
    \end{subfigure}
    \hfill
    \begin{subfigure}[b]{0.49\linewidth}
        \centering
        \includegraphics[width=\textwidth]{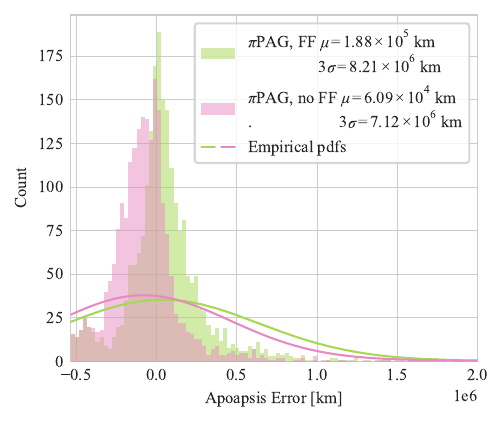}
        \caption{$\pi$PAG.}
    \end{subfigure}
    \caption{Captured apoapsis error histogram comparing \piPAG and \eFNPAG with and without a fading filter for density estimation for Gaussian near-escape conditions.}
    \label{fig:apo_hist_escape_ff}
\end{figure}

Additionally, the fading filter statistically significantly improves apoapsis targeting for all pairwise comparisons between guidance with a fading filter and without a fading filter. Histograms of apoapsis mean error for both guidance algorithms with and without the fading filter are shown in \cref{fig:apo_hist_escape_ff}. The $p$ values for comparisons between (1) \eFNPAG and \eFNPAG without a fading filter and (2) \piPAG and \piPAG without a fading filter for both scenarios are 0.000. 

 This indicates statistically significant decrease in apoapsis error mean and standard deviation for both \eFNPAG and \piPAG when using the fading filter in the near-impact scenario. However, this also indicates that the reduction in apoapsis error when using \piPAG without a fading filter in the near-escape scenario is significant. As \cref{tab:ff_performance} presents statistics for captured trajectories only, and \piPAG without the fading filter captures fewer cases, the reduced error is expected. As \piPAG with the fading filter successfully recovers a larger number of trajectories, including more extreme cases, the mean apoapsis error of its captured cases is higher than for \piPAG without the fading filter, which fails to recover these extreme trajectories.

\subsection{Implications of GMVAE Selection on \piPAG Performance}
\label{sec:alt_gmvae}

In \cref{sec:hyperparameter}, GMVAE accuracy was compared in terms of weighted misassignment rate (\cref{eqn:weighted_rate}) and failure-only weighted misassignment rate (\cref{eqn:failure_weight}). The preceding sections used the best-performing GMVAE in terms of weighted misassignment rate. In this section, we will compare \piPAG using the best weighted misassignment rate GMVAE to \piPAG using the best failure-only weighted misassignment rate GMVAE, hereafter referred to as f\piPAG.

A 2,500-sample Monte Carlo simulation was run for the near-escape and near-impact scenario using $\varepsilon$FNPAG, \piPAG, and f$\pi$PAG. The samples were drawn from the distributions described in \cref{tab:diffdist_samedp_dispersions} using LHS, and the truth atmosphere model is GRAM with a \texttt{dp} = 2. 

\cref{tab:diff_gmvae_outcome} compares outcomes for each scenario. For the near-escape scenario, f\piPAG saves more recoverable escape trajectories than $\pi$PAG, reducing the overall failure rate. For the near-impact scenario, there is no difference in capture rate between \piPAG and f$\pi$PAG. Both algorithms salvage all recoverable impacts. For the near-impact scenario, the escape rate increases, as one previously captured case escapes (see discussion in \cref{sec:diffdist_samedp}). 

\begin{table}[htb]
    \centering 
    \caption{\piPAG and \eFNPAG outcomes for different GMVAEs.}
    \label{tab:diff_gmvae_outcome}
	\begin{tabular}{p{3.5cm}lccc>{\centering\arraybackslash}p{2cm}>{\centering\arraybackslash}p{2cm}}
	\hline
	Scenario & Guidance & Capture \% & Impact \% & Escape \% & Recoverable Escape \% & Recoverable Impact \% \\
	\hline
	\multirow{3}{=}{Gaussian Near-Escape (\texttt{dp} = 2)} & $\varepsilon$FNPAG & 84.84 & 0.00 & 15.16 & 0.00 & N/A \\
	 & $\pi$PAG & 86.64 & 0.00 & 13.36 & 75.41 & N/A \\
	 & f$\pi$PAG & 86.96 & 0.00 & 13.04 & 86.89 & N/A \\
	\hline
	\multirow{3}{=}{Gaussian Near-Impact (\texttt{dp} = 2)} & $\varepsilon$FNPAG & 95.04 & 4.92 & 0.04 & N/A & 0.00 \\
	 & $\pi$PAG & 95.76 & 4.16 & 0.08 & N/A & 100.00 \\
	 & f$\pi$PAG & 95.76 & 4.16 & 0.08 & N/A & 100.00 \\
	\hline
	\end{tabular}
\end{table}

The apoapsis and $\Delta v$ requirements for each guidance algorithm are shown in \cref{tab:diff_gmvae_perf}. The difference in apoapsis error for \eFNPAG and \piPAG is not statistically significant, however f\piPAG statistically significantly improves apoapsis mean error and increases standard deviation over \eFNPAG with a $p$ value of zero in a Mann-Whitney U test. Both versions of \piPAG statistically significantly increase mean total $\Delta v$ and decrease spread over \eFNPAG with a $p$ value of zero for each.

\begin{table}[htb]
    \centering 
    \caption{\piPAG and \eFNPAG you uFNPAG performance comparison for different GMVAEs.}
    \label{tab:diff_gmvae_perf}
\begin{tabular}{p{3.5cm}lcccc}
\hline
Scenario & Guidance & \begin{tabular}{@{}c@{}} $r_a$ Mean \\ $[\times 10^{4}$ km$]$\end{tabular} & \begin{tabular}{@{}c@{}}$r_a$ 3$\sigma$ \\ $[\times 10^{5}$ km$]$ \end{tabular} & \begin{tabular}{@{}c@{}}$\Delta v$ Mean  \\ $[\times 10^{2}$ m/s$]$ \end{tabular} & \begin{tabular}{@{}c@{}}$\Delta v$ 3$\sigma$ \\ $[\times 10^{3}$ m/s$]$ \end{tabular} \\
\hline
\multirow{3}{=}{Gaussian Near-Escape (\texttt{dp} = 2)} & $\varepsilon$FNPAG & 15.73 & 41.15 & 2.97 & 1.41 \\
 & $\pi$PAG & 18.78 & 82.12 & 3.42 & 1.36 \\
 & f$\pi$PAG & 7.99 & 71.08 & 3.48 & 1.32 \\
\hline
\multirow{3}{=}{Gaussian Near-Impact (\texttt{dp} = 2)} & $\varepsilon$FNPAG & -2.71 & 0.99 & 10.68 & 2.89 \\
 & $\pi$PAG & -2.39 & 1.49 & 12.13 & 2.71 \\
 & f$\pi$PAG & -1.66 & 1.46 & 12.50 & 2.57 \\
\hline
\end{tabular}
\end{table}

\cref{tab:diff_gmvae_activation} describes the cases in which a corrective guidance action was taken based on the GMVAE predicted modal probabilities. This indicates that either capture probability fell below $\epsilon_C$ or failure (escape or impact) probability exceeded $\epsilon_F$. Corrective guidance is activated in more cases when using the GMVAE with the best failure-only weighted misassignment rate (f$\pi$PAG); for the near-impact scenario, corrective action is taken for all 2,500 trajectories in f$\pi$PAG. The number spurious corrective guidance activations, indicated in the ``Prev. Captured'' column, increases when using f$\pi$PAG. However, corrective guidance is taken for all 61 recoverable escape trajectories when using f\piPAG versus 57 for $\pi$PAG. 

\begin{table}[htb]
    \centering 
    \caption{\piPAG activation statistics for different GMVAEs.}
    \label{tab:diff_gmvae_activation}
\begin{tabular}{p{3.5cm}lcccc>{\centering\arraybackslash}p{1.3cm}>{\centering\arraybackslash}p{1cm}>{\centering\arraybackslash}p{1cm}}
\hline
Scenario & Guidance & Total & Captured & Escaped & Impact & Prev. Captured & Rec. Escape & Rec. Impact \\
\hline
\multirow{2}{=}{Gaussian Near-Escape (\texttt{dp} = 2)} & $\pi$PAG & 657 & 326 & 331 & 0 & 281 & 57 & N/A \\
 & f$\pi$PAG & 1222 & 896 & 326 & 0 & 843 & 61 & N/A \\
\hline
\multirow{2}{=}{Gaussian Near-Impact (\texttt{dp} = 2)} & $\pi$PAG & 429 & 323 & 2 & 104 & 304 & N/A & 19 \\
 & f$\pi$PAG & 2500 & 2394 & 2 & 104 & 2375 & N/A & 19 \\
\hline
\end{tabular}
\end{table}

\section{Discussion}
\label{sec:discussion}

Aerocapture is a mission-critical, single-opportunity maneuver, making even marginal improvements in capture success rate consequential for mission viability. As shown in the results in \cref{sec:results}, $\pi$PAG, which utilized corrective guidance actions informed by the probabilistic indicator function, consistently improved capture success rates compared with those of the deterministic guidance algorithm $\varepsilon$FNPAG in all scenarios. Some aerocapture GNC Monte Carlo studies produce multiple failures for large sample sets~\cite{rataczak_convex_nodate,deshmukh_performance_2024}; thus, reducing even a small number of failures in large Monte Carlo studies meaningfully increases confidence in aerocapture as a viable orbit insertion technique for outer planet missions.

A key strength of the GMVAE as a probabilistic indicator is that it achieves 100\% recovery of all recoverable cases within its training conditions, demonstrating that a well-matched training distribution enables near-perfect classification. The GMVAE is also capable of extrapolating to conditions outside its training data. Notably, for cases outside the training conditions, \piPAG saves all recoverable impact trajectories, while recovering between 71.43\% and 86.89\% of recoverable escaped trajectories. The asymmetry between impact and escape recovery reflects the conservative bias of the selected GMVAE: by overpredicting impacts, the classifier reliably intervenes for near-impact trajectories but may not flag marginal escape cases. While this overprediction improves \piPAG success rates in near-impact cases, it also applies more spurious corrections. This tradeoff is most apparent in the near-impact datasets where \piPAG converts previously captured cases to escapes: a conservative probabilistic indicator strategy can result in losing some previously captured trajectories. 

\begin{figure}[htbp]
    \centering
    \begin{subfigure}[b]{0.49\linewidth}
        \centering
        \includegraphics[width=\textwidth]{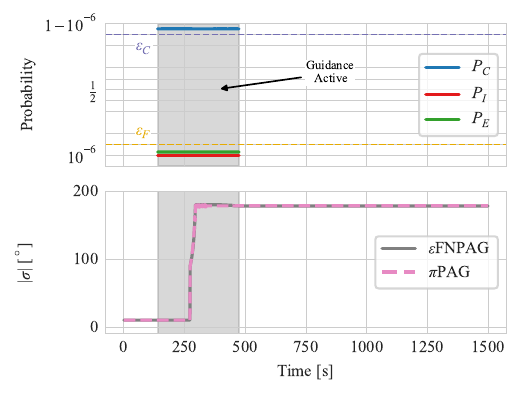}
        \caption{\piPAG escaped case where GMVAE is not activated.}
        \label{fig:gmvae_not_act}
    \end{subfigure}
    \hfill
    \begin{subfigure}[b]{0.49\linewidth}
        \centering
        \includegraphics[width=\textwidth]{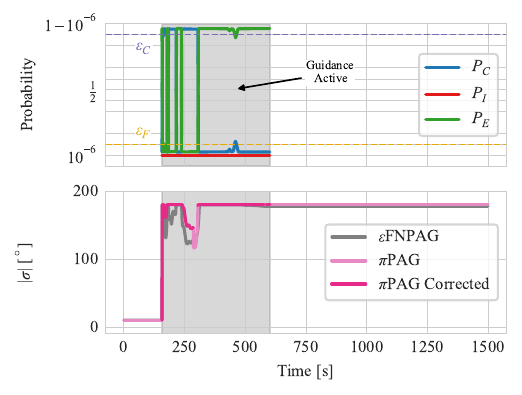}
        \caption{\piPAG escaped case where GMVAE is activated.}
        \label{fig:gmvae_act}
    \end{subfigure}
    \caption{Examples of \piPAG failed recoveries.}
    \label{fig:piPAG_gmvae_fails}
\end{figure}

Two distinct failure mechanisms account for the cases \piPAG cannot recover, illustrated in \cref{fig:piPAG_gmvae_fails}. For these initial condition, vehicle, and atmosphere dispersions, both \eFNPAG and \piPAG result in escape. In the first, the GMVAE never identifies a high failure probability; the trajectory appears captured to the classifier throughout, and no correction is triggered (\cref{fig:gmvae_not_act}). In the second, corrective guidance is applied but the bank angle correction magnitude and persistence time are insufficient to redirect the trajectory before atmospheric exit (\cref{fig:gmvae_act}). Distinguishing between these mechanisms is important when selecting a GMVAE and tuning $\pi$PAG. To avoid the first failure mode, a more accurate GMVAE for cases outside its training data is required. To avoid the second failure mode, \piPAG parameters could be tuned to result in a captured trajectory. This demonstrates that although \piPAG is a robust algorithm which can improve capture success rate, care must be taken to select both a GMVAE that can classify cases outside its training data and corrective guidance parameters to enable mission success. 

The increase in required $\Delta v$ when using \piPAG reflects a key tradeoff when increasing robustness through a probabilistic indicator function: while capture rate increases, required $\Delta v$ does as well. This is due to both the recovery of highly elliptical trajectories that inherently require more post-capture correction and spurious corrections applied to trajectories that would have captured without intervention. For problems with high uncertainty where robustness is desired, \piPAG recovers more cases. However, for more fuel-constrained problems or scenarios with lower uncertainty, \eFNPAG produces more favorable behavior. This effect is also seen when investigating GMVAE selection, as employing a more conservative GMVAE results in a higher success rate at the cost of higher propellant usage.

A different tradeoff is present when comparing the performance of the probabilistic indicator to a fading filter for density estimation in NPC aerocapture guidance. Because \piPAG without a fading filter does not improve over \eFNPAG with a fading filter with regards to apoapsis error or capture rate, a probabilistic indicator function is not a pure replacement for a first-order fading filter. The fading filter improves apoapsis error by accounting for the differences between the truth and onboard atmosphere models through incorporating state measurements. While the fading filter improves apoapsis radius error variability, \piPAG improves the capture rate by recovering failure cases. Thus, these two tools can be used in conjunction to improve aerocapture performance for high-uncertainty scenarios due to their complementary benefits.

GMVAE selection also modulates the robustness–propellant usage tradeoff: a classifier selected based on overall accuracy and one optimized for failure detection produce meaningfully different guidance behaviors. The baseline GMVAE, which was superior in terms of overall weighted misassignment rate, resulted in fewer recovered trajectories but required lower propellant usage, while the alternative GMVAE, which was superior in terms of failure-only weighted misassignment rate, recovered more escaped trajectories at the expense of more $\Delta v$. This indicates that to prioritize safety, it is better to select a GMVAE that accurately predicts all failures, while misclassifying some captures, than prioritizing a GMVAE with a perfect capture prediction rate. 

While \piPAG demonstrated robust performance across the scenarios tested, several improvements could further extend its applicability. The results indicate that while the GMVAE is an effective classifier for cases within its training conditions and extrapolates well to conditions outside its training conditions, the GMVAE applicability is  also limited by flight scenario and destination. For example, the GMVAEs presented in this work could not be applied to aerocapture on another planet or for significantly different entry interface conditions at Uranus. GMVAEs trained for other entry conditions or destinations are expected to demonstrate similar performance improvements. Additionally, a key weakness of this implementation of \piPAG is inaccurate escape classification outside the training data. More GMVAE hyperparameter sweeps could be conducted to improve GMVAE prediction accuracy for escape cases, or  the regularization probability in GMVAE training could be reduced to further decrease the failure probability threshold. Additionally, efforts could be made to improve the performance of a probabilistic indicator function through curriculum learning~\cite{rataczak_density_nodate}. Further iterations of the GMVAE could be trained on \piPAG simulations rather than \eFNPAG simulations so that the GMVAE training data includes corrective guidance actions. Finally, in this work, probabilistic indicator functions were only applied to lateral maneuvers for a lift-modulation in $\varepsilon$FNPAG. However, corrective action could be taken for other entry control schemes---such as direct force control~\cite{engel_assessment_2025,engel_optimal_2025} or drag modulation~\cite{putnam_drag-modulation_2014}---or in other aerocapture predictor-correctors~\cite{cihan_analytical_2021,chen_augmented_2024}.

\section{Conclusion}

Aerocapture\textemdash particularly when targeting outer planets or operating under high uncertainty and limited navigation precision\textemdash faces significant risks due to trajectory dispersions and environmental variability. Traditional guidance algorithms typically rely on the mean state for decision-making and do not account for the likelihood of failure modes. This can result in mission failure when the true state deviates from onboard predictions, affecting both low-cost missions with large uncertainties and missions with precise navigation solutions. The probabilistic indicator function approach presented here improves upon this by estimating the likelihood of each terminal mode and incorporating that information into the guidance logic. This risk-aware strategy increases the mission success rate by proactively biasing guidance commands away from failure regions, making it a critical advancement for robust autonomous aerocapture in uncertain environments. The method is computationally efficient and straightforward to integrate with existing numeric predictor-corrector algorithms, making it an attractive fault-tolerance addition for guided aerocapture missions.

In this work, a novel probabilistic indicator function predictor-corrector aerocapture guidance algorithm (\piPAG) was developed and shown to improve aerocapture success rates under high uncertainty. The algorithm uses a Gaussian Mixture Variational Autoencoder (GMVAE) to predict the probability of capture or escape for a given trajectory, and biases the guidance bank angle commands to increase the chance of capture. The algorithm was shown to be effective in recovering both escape and impact cases, which would otherwise fail with the state-of-the-art Fully Numeric Predictor-Corrector Aerocapture Guidance algorithm. The results focused on four scenarios. The first analysis demonstrated that probabilistic corrective guidance actions can be taken to save both impact and escape cases. The second study showed that \piPAG can be effective even when the data used for GMVAE training are not representative of the true environment. In the third study, we compared the use of a probabilistic indicator function to a first-order fading filter for density estimation and found that each of these mechanisms provide complementary performance benefits. A probabilistic indicator function is not a replacement for improved density estimation, as the former improves capture success rate while the latter improves apoapsis targeting accuracy (and thus required $\Delta v$). Finally, the best failure-only weighted misassignment rate GMVAE improved capture success rate over the best weighted misassignment rate GMVAE, at the expense of higher $\Delta v$ due to spurious corrective guidance activations and the recovery of highly-elliptical cases. All results show that a simple probabilistic indicator evaluation to inform a corrective guidance action can increase aerocapture robustness at minimal computational cost. 

\appendix
\section{Numeric Predictor-Corrector Objective Function Performance in FNPAG}
\label{app:fnpag_obj}

Rataczak et al. introduced a novel objective function for aerocapture guidance based on energy rather than apoapsis radius~\cite{rataczak_convex_nodate}. This function is desirable over the original apoapsis radius expression presented by Lu et al.~\cite{lu_optimal_2015}:
\begin{equation}
	r_a(r_{\text{exit}}, V_{\text{exit}}, \gamma_{\text{exit}}) - r_a^* = 0,
\end{equation}
as it has no singularities and monotonically decreases throughout the trajectory. While the results in Rataczak et al.'s paper demonstrated that this expression improved apoapsis targeting in the presence of $J_2$, here we demonstrate that using energy expression as the NPC objective, even when not considering $J_2$ effects, improves FNPAG's desirable outcomes. FNPAG using the energy expression in \cref{eqn:energy} is denoted as $\varepsilon$FNPAG to distinguish from the original FNPAG formulation using apoapsis radius in the NPC objective. 

The algorithms were evaluated via a 2,500-sample Monte Carlo using the near-escape Gaussian initial conditions described in \cref{tab:diffdist_samedp_dispersions} sampled with LHS. The truth atmosphere were samples from GRAM with a \texttt{dp} = 2. The same initial condition and vehicle parameter samples were evaluated using FNPAG and $\varepsilon$FNPAG. \cref{tab:fnpag_obj_outcome} demonstrates that the capture rate decreases substantially when using the original NPC objective function, while using the energy expression results in more captures. 

\begin{table}[hbt]
    \centering 
    \caption{Comparison of outcomes when using different NPC objective functions in FNPAG.}
    \label{tab:fnpag_obj_outcome}
    \begin{tabular}{lccc}
	\hline
	Guidance & Capture [\%] & Impact [\%] & Escape [\%] \\
	\hline
	FNPAG & 67.04 & 0.00 & 32.96 \\
	$\varepsilon$FNPAG & 84.84 & 0.00 & 15.16 \\
	\hline
	\end{tabular}
\end{table}

The alternative objective function also improved the mean and spread of apoapsis error and reduced required propellant, as seen in \cref{tab:fnpag_obj_perf}. 
\begin{table}[hbt]
    \centering 
    \caption{Comparison of performance when using different NPC objective functions in FNPAG.}
    \label{tab:fnpag_obj_perf}
	\begin{tabular}{lcccc}
	\hline
	 Guidance & \begin{tabular}{@{}c@{}} $r_a$ Mean \\ $[\times 10^{4}$ km$]$\end{tabular} & \begin{tabular}{@{}c@{}}$r_a$ 3$\sigma$ \\ $[\times 10^{5}$ km$]$ \end{tabular} & \begin{tabular}{@{}c@{}}$\Delta v$ Mean  \\ $[\times 10^{2}$ m/s$]$ \end{tabular} & \begin{tabular}{@{}c@{}}$\Delta v$ 3$\sigma$ \\ $[\times 10^{3}$ m/s$]$ \end{tabular} \\
	\hline
	FNPAG & 38.93 & 180.39 & 3.44 & 1.57 \\
	$\varepsilon$FNPAG & 15.73 & 41.15 & 2.97 & 1.41 \\
	\hline
	\end{tabular}
\end{table}

Thus, the authors selected the energy expression for NPC to ensure \piPAG is compared with an improved version of FNPAG for the conditions investigated here.

\section*{Funding Sources}
This work was supported by NASA Space Grant Technology Research Fellowship grant number 80NSSC23K1227. AD's work is supported by the DOE-NNSA under award number DE-NA0003968 and NASA under award number 80NSSC21K1117.

\section*{Acknowledgments}
We acknowledge Alexandre Cortiella for providing an early version of the GMVAE codes; we thank Jens Rataczak for his help implementing FNPAG and GRAM and guiding discussions on this algorithm; and we appreciate Chun-Wei Kong for his assistance with the GMVAE training architecture setup. Artificial intelligence tools were used for \LaTeX~ equation editing, Python coding assistance in figure generation, and language editing to improve clarity and readability.

\bibliography{references}

\end{document}